\documentclass[useAMS,usenatbib]{mn2e}
\usepackage{aas_macros}
 \usepackage{epsfig}
 \usepackage{stfloats}
 \usepackage{relsize}
 \usepackage{fixltx2e}
 \usepackage{josty}
 \usepackage{graphicx}
\usepackage[usenames,dvipsnames]{color}

\newcommand{\sfc}[1]{\textcolor{Black}{{ #1}}} 
\newcommand{\arxc}[1]{\textcolor{Black}{{ #1}}}
\newcommand{\refc}[1]{\textcolor{Black}{{ #1}}}
\newcommand{\jwc}[1]{\textcolor{Black}{{ #1}}}

\title[Quenching: Compact Centres and Massive Haloes]
{Two Conditions for Galaxy Quenching: Compact Centres and Massive Haloes}

\author[Woo \etal] 
    {\parbox{\textwidth}{Joanna
        Woo$^{1,2}$\thanks{joaw@phys.huji.ac.il, joanna.woo@phys.ethz.ch}, 
Avishai
    Dekel$^{2}$, S. M. Faber$^{3}$, David C. Koo$^{3}$}
\vspace{0.4cm}\\
\parbox{\textwidth}{ $^{1}$Institute for Astrophysics, Department of
  Physics, ETH Zurich,
  Wolfgang-Pauli-Strasse 27, CH-8093 Zurich, Switzerland\\
  $^{2}$Center for Astrophysics and Planetary Science, Racah Institute
  of Physics, The Hebrew University, Jerusalem 91904, Israel\\
  $^{3}$University of California Observatoires/Lick Observatory,
  Department of Astronomy and Astrophysics, University of California, 
  Santa Cruz, CA 95064, USA\\
}}

\pagerange{\pageref{firstpage}--\pageref{lastpage}} \pubyear{2012}

\begin{document}
\label{firstpage}
\maketitle

\begin{abstract}
  We investigate the roles of two classes of quenching mechanisms 
  \sfc{for central and satellite galaxies in the SDSS ($z<0.075$)}: those
  involving the halo and those
  involving the formation of a compact centre.  
  For central galaxies with inner compactness $\sigone \sim 10^{9-9.4}\Msun \kpc^{-2}$, 
  the quenched fraction $\fq$ is strongly correlated with $\sigone$ with only weak
  halo mass $\Mh$ dependence.  However, at higher and lower $\sigone$, sSFR is a
  strong function of $\Mh$ and mostly independent of $\sigone$.
  \sfc{In other words, $\sigone \sim 10^{9-9.4} \Msun \pkpc2$ divides galaxies into 
  those with high sSFR below and low sSFR above this range.  
  In both the upper and lower regimes, increasing $\Mh$ shifts the entire sSFR 
  distribtuion to lower sSFR without a qualitative change in shape.  This is true even at 
  fixed $\Ms$, but varying $\Ms$ at fixed $\Mh$ adds no quenching information.}
  Most of the quenched centrals with $\Mh > 10^{11.8}\Msun$ are 
  dense ($\sigone > 10^{9}~ \Msun \kpc^{-2}$),
  suggesting compaction-related quenching maintained by halo-related
  quenching.  However, 21\% are diffuse, indicating only
  halo quenching.  
  For satellite galaxies in
  the outskirts of halos, quenching is a strong function of
  compactness and a weak function of host $\Mh$.  
  In the inner halo, $\Mh$ dominates quenching, with
  $\sim 90\%$ of the satellites being quenched once $\Mh >
  10^{13}\msun$.  \refc{This regional effect is greatest for the least massive satellites.}
  As demonstrated via
  semi-analytic modelling with simple prescriptions for quenching, the
  observed correlations can be explained if quenching due to central
  compactness is rapid while quenching due to halo mass is slow.
\end{abstract}

\begin{keywords}
galaxies: evolution -- star formation -- haloes -- groups -- structure
\end{keywords}

\section{Introduction}
\label{intro}

Large surveys of galaxies over the last decade have established that
the nature of galaxies is bimodal.  Their populations are effectively
described as either blue and star-forming or red and quiescent.  The
star-forming population tends to exhibit disk-like morphologies and
low central densities, while the quiescent galaxies host a dominant
bulge component with high central densities
\citep{str01,kau03a,bla03,bald04,wuyts11,cheung12}.
While galaxies are expected to accrete gas and form stars, it is still
a mystery how the quiescent population originated.  

Several mechanisms have been proposed to explain the shutdown of star
formation.  Among them, the ``halo quenching'' mechanism has been the
major reason behind the success of galaxy formation models in
reproducing several aspects of the bimodality
\citep{croton06,bow06,cattaneo08}.  Quenching \sfc{in these models} hinges upon 
a critical
halo mass $\Mcrit$ (a few $\times 10^{12}\:\Msun$), below which
accretion is cold and conducive to star formation.  Above $\Mcrit$,
infalling gas reaches the sound speed and a stable shock forms.  This
shock heats the gas and prevents its accretion onto the central galaxy
to form stars \citep{ree77,bir03,ker05}.  For a population of haloes,
this should not be interpreted as a sharp threshold but rather as a
mass range extending over one or two decades where there is a decrease
in the cold accretion as a function of halo mass ($\Mh$)
\citep{ocv08,keres09,vandevoort11}.

While halo quenching successfully reproduces the bimodality of star
formation \refc{when implemented as a strong, immediate effect}, it does not 
naturally explain the link between quiescence
and a prominent bulge, structural compactness, \sfc{or high central density}.
Among the proposed bulge-building/compacting quenching mechanisms is the major
merger scenario, in which a central spheroid forms through the violent
relaxation of pre-merger stars \citep{toomre72,mihos94,hopkins09}.
Furthermore, gas is driven to the central regions of the system,
producing a starburst that quickly exhausts the gas (through
consumption and winds) and contributes to the central compactness.  AGN
triggered in the merger also suppress star formation.  However, it is
uncertain whether major mergers actually quench galaxies
\citep{coil11}, or whether the merger rate is sufficient to account
for the number of bulges \citep{hopkins10,lotz11}.

Another process that builds a compact bulge is gaseous inflow through violent
disk instability, including the migration of giant star-forming clumps
\citep{noguchi99,bournaud07,dekel09,mandelker14,dekel14}.  Infalling
cold streams maintain the unstable disk and replenish the gas.  These
gaseous inflows may fuel AGN and consume gas via high star-formation
rates and powerful outflows creating a central concentration of young
stars in a dense ``blue nugget''.  This process depends on a strong
gas inflow rate, which occurs at $z\gtsima 2$, but declines with time.

The build-up of the bulge through mergers or instabilities can
stabilise the disc against further star formation (see also
\citealp{martig09}).  This can also be coupled with a mechanism that
shuts off accretion, such as virial shock heating or AGN heating,
which cuts the supply of the gas that fuels both star formation and
disc instabilities.

As for satellite galaxies, the above mechanisms may have operated on
these while they were still centrals.  However, there are additional
quenching mechanisms that are unique to satellites such as
strangulation \refc{(the cut off cold accretion -} 
\citealp{larson80,balogh00}), 
ram pressure stripping \refc{(the stripping of both cold and hot gas due to 
movement
through a hot medium -} 
\citealp{gunn72,abadi99}), tidal stripping \refc{(the stripping of gas and 
stars outside the
the tidal radius as a satellite approaches pericentre -} \citealp{read06}) and 
harassment
\refc{(heating of cold gas due to high-speed satellite-satellite interactions 
-} \citealp{moore98,villalobos12}).  
These mechanisms are expected to be
more efficient near the centres of haloes.

Several studies have attempted to observationally constrain or rule
out any of these scenarios with conflicting results.  For example,
quenching for central galaxies is observed to correlate more strongly
with an estimate of halo mass $\Mh$ than with luminosity
\citep{weinmann06} or stellar mass $\Ms$ \citep{woo13}, supporting the
halo quenching scenario (see also \citealp{tal14}).  
Halo quenching
is also consistent with the many studies that find a dependence of
quenching on environment and clustering 
\citep{hog03,kau04,bal04,blanton05b,baldry06,bundy06,coo08,
skibba09a,wilman10,quadri12,haas12,hartley13}.
However,
quenching also has a strong dependence on morphology and galaxy
structure in the local universe
\citep{kau03,franx08,bell08,vandokkum11,robaina12,omand14,bluck14} and
at high-$z$ \citep{wuyts11,cheung12,bell12,szomoru12,wuyts12,barro13,lang14},
which is not expected in halo quenching, but in bulge-building
mechanisms.  Furthermore, galaxies which may be transitioning from
star formation to quiescence appear to have early type morphologies
\citep{mendel13}.  On the other hand, while bulge-building mechanisms
succeed in producing compact inner regions, they do not address the
observed quenching of disk galaxies \refc{which may represent 25-65\% of 
quenched galaxies, 
especially at high-$z$}
(\citealp{stockton04,mcgrath08,vanD08,vandenbergh09,bundy10,vanderwel11,salim12,
bruce12,bruce14}).  \sfc{The quenching of such galaxies is more 
naturally 
explained as a 
halo process, whether as centrals \citep{vandenbergh09,bundy10} or satellites 
\citep{bell12,peng10,peng12,knobel13,kovac14}}.  Thus it remains
unclear how important the halo environment is compared to galaxy
morphology/structure in predicting quenching.

For satellites, the quenching picture is not any clearer.  Quenching
has been observed to correlate with cluster/group-centric distance
\citep{gomez03,bal04,tanaka04,rines05,blanton07,haines07,wolf09,
  hansen09,vonderlinden10,woo13} supporting the various satellite
quenching scenarios that operate in dense environments.  
The quenching
of satellites also seems to depend on environment more than morphology
\citep{koopmann98,goto03,vogt04,kodama04,wolf09,bamford09}.  This can
naturally be explained by the
presence of hot gas in massive haloes \citep{gabor14}, which is
conducive to, for instance, ram
pressure stripping.  However, others have argued that colours and red
fractions of satellites are determined by a galaxy's stellar mass
$\Ms$ \citep{vandenbosch08,peng10,peng12} and number density of
surrounding galaxies \citep{peng10,peng12} rather than by an estimate
of its host halo mass $\Mh$.  \cite{woo13}, on the other hand, showed
that the importance of a satellite's $\Ms$ vs. its host $\Mh$ in
predicting quenching depends on where the satellite lies in its halo.
This could reflect residual ``central'' quenching before infall,
coupled with a ``delayed-then-rapid'' form of satellite quenching
\citep{wetzel13,trinh13,mok13}.  \cite{omand14} recently showed that
satellite structure rather than $\Ms$ determines satellite quenching,
but \cite{carollo14} showed that the fraction of morphologically
early-type satellites is independent of environment.  Some of these
confusions can be resolved by comparing the importance of
bulge-related processes on satellites to halo related
processes at different group-centric radii.

The goal of this paper is to study the effectiveness of halo-related
and bulge/compactness-related quenching in centrals and satellites
in the SDSS.  

``Compactness-related'' quenching (also related to the bulge) will be
measured by the central surface density within the inner 1 kpc
($\sigone$) following \cite{cheung12,fang13} since this quantity
probes quenching mechanisms that involve gaseous inflows toward a
galaxy's central regions.  Additionally, \cite{cheung12,fang13} showed
that $\sigone$ is a stronger predictor for quenching than the
bulge-to-total ratio $B/T$ and S\'{e}rsic $n$.

``Halo-related'' quenching for centrals can in principle be explored
using direct estimates of $\Mh$ or using $\Ms$ as a proxy since the
two are tightly related for centrals.  
\sfc{In fact, \cite{fang13}
showed that the combination of the inner compactness and stellar mass $\Ms$ 
strongly predicts quenching for central galaxies.  
Since \cite{woo13} showed that $\Mh$ 
predicts quenching for centrals better than $\Ms$, this motivates a 
similar analysis with $\Mh$ instead of $\Ms$.  Moreover, $\Mh$ provides some 
advantages
over $\Ms$.  First, $\Ms$ confuses the relation between internal and external 
quenching
mechanisms since it is an internal property which tightly correlates with
the inner compactness, as well as with $\Mh$ for centrals.  
Thus, it is unclear whether correlations of quenching with $\Ms$ point to 
halo-related 
quenching or to internal quenching processes.
Second, as described above, 
there are quenching 
mechanisms that are directly related to the halo and to the inner compactness, 
but there are no known mechanisms for quenching that are directly related to
the stellar mass for massive galaxies.  The net result of this
analysis will validate the use of $\Mh$ over $\Ms$ for centrals as expected
theoretically.  Lastly, $\Ms$ satellites is largely unrelated to host $\Mh$ and 
any associated halo processes.}

Quenching will be measured using the quenched fraction
$\fq$ of galaxies along with the sSFR = SFR/$\Ms$.  These are defined
in \sec{data} along with all data used in this study.  Our results for
centrals will be presented in \sec{centralssection} and satellites in
\sec{satellites}.  \sec{discussion} will discuss an interpretation of
our findings as a result of slow and rapid quenching, with concluding
remarks in \sec{conclusion}.

\jwc{This analysis assumes concordance cosmology: $H_o = 70~{\rm km~s^{-1}
  Mpc^{-1}}, \Omega_M = 0.3, \Omega_\Lambda=0.7$.  Halo masses are converted to this 
cosmology with $\sigma_8=0.9, \Omega_b =0.04$.}

Our conclusions from this analysis will be that both halo-related and
compactness-related quenching govern the evolution of centrals and
satellites in such a way that may point to a difference in duration
for the two types of quenching.

\section{Data}
\label{data}

\subsection{The sample}
\label{sdsssample}

The SDSS \citep{york00,gunn06}
sample used throughout this analysis is from the Data Release 
7 \citep{abazajian09}
limited to the redshift range to $0.005 < z < 0.075$.  This sample
contains 65 939 
galaxies after matching all catalogues and applying
all cuts as described below.

Using the K-correction utilities of \cite{blanton07} (v4\_2), we
calculate $V_{\rm max}$ from $ugriz$ photometry (``petro'' values - \citealp{gunn98,doi10}) and
redshifts from the NYU-VAGC (DR7) catalogue
\citep{blanton05,adelman08,padmanabhan08} and the $r$-band limit
(17.77) of the spectroscopic survey.  We weight each galaxy by its
$1/V_{\rm max}$ multiplied by the inverse of its spectroscopic
completeness (also obtained from the NYU-VAGC).  All quoted galaxy
fractions and volume densities are weighted.

The NYU-VAGC contains 2 506 754 objects, of which 207 005 
are classified as galaxies ({\tt SpecClass}=2), is a primary spectroscopic
object ({\tt specprimary}=1), is an object for which {\tt KCORRECT} produced finite
$V_{\rm max}$, and lies within our chosen redshift range.

\subsection{Stellar masses}
\label{mssdss}
Stellar mass $\Ms$ is taken from the DR7 catalogue provided online by
J. Brinchmann et
al.\footnote{http://www.mpa-garching.mpg.de/SDSS/DR7/\label{brinchsite}}.
Using a \cite{kro01} IMF, they derived $\Ms$ through SED fitting
similar to the method that \cite{sal07} used for estimating SFRs, and
the method used by \cite{kau03a} to derive $\Ms$ by fitting spectral
features rather than photometry.  These stellar mass estimates differ
from \cite{kau03a} by less than 0.1 dex for $\Ms \gtsima 10^{9}
\Msun$.  The formal $1$-$\sigma$ errors are typically about 0.05 dex
or less (from the 95$\%$ confidence intervals of the probability
distribution).

This catalogue contains 927 552 objects, 198 159 
of which match the
NYU-VAGC and pass our cuts described above.  Of these, 182 406 
are above our cut of $\Ms = 10^{9} \Msun$.

\subsection{SFR} 
\label{sdsssfr}

\cite{bri04} estimated SFR for the DR4, and those used here are an
updated and improved version for the DR7.  They are provided online by
J. Brinchmann et al.$^{\ref{brinchsite}}$.  As in \cite{bri04}, these
SFR estimates are calibrated to the \cite{kro01} IMF, and combine
measurements inside and outside the fibre.  Inside the fibre, SFR is
estimated from $\Ha$ and $\Hb$ lines for star-forming galaxies.  SFR for
galaxies with weak lines or showing evidence for AGN is estimated
using the relation between the D4000 break and $\Ha$ and $\Hb$ lines
observed in star-forming galaxies.  Outside the fibre, they fit
stellar population models to the observed photometry following the
method of \cite{sal07}.  The means of the resulting probability
distribution function of SFR were added to the SFR estimates in the
fibre for an estimate of the total SFR.
\refc{These SFR estimates account for dust in the SED modelling, and we have 
confirmed that applying a selection on inclination ($b/a > 0.5$) does not 
affect our results. }

The original catalogue contains 927 552 objects, 182 406 
of which pass
our previous cuts.  Of these, we select 182 125
that are flagged with {\tt sfflag} = 0 (all other SFR's
were measured with slightly different methods and could introduce
bias).

The typical errors of these SFRs are estimated by J. Brinchmann
(private communication) to be about 0.4 dex for star-forming galaxies,
0.7 dex for intermediate galaxies and 1 dex or more for dead galaxies.
For the dead galaxies, the given SFR values should be considered as upper limits rather than measurements.
However, a
comparison of these SFR estimates with those of \cite{sal07}, which
are derived from SED fitting of UV and optical light from GALEX,
suggest that the error is closer to 0.2 dex for star-forming galaxies
(K. Bundy, private communication).

\begin{figure}
\epsscale{1.0}
\plotone{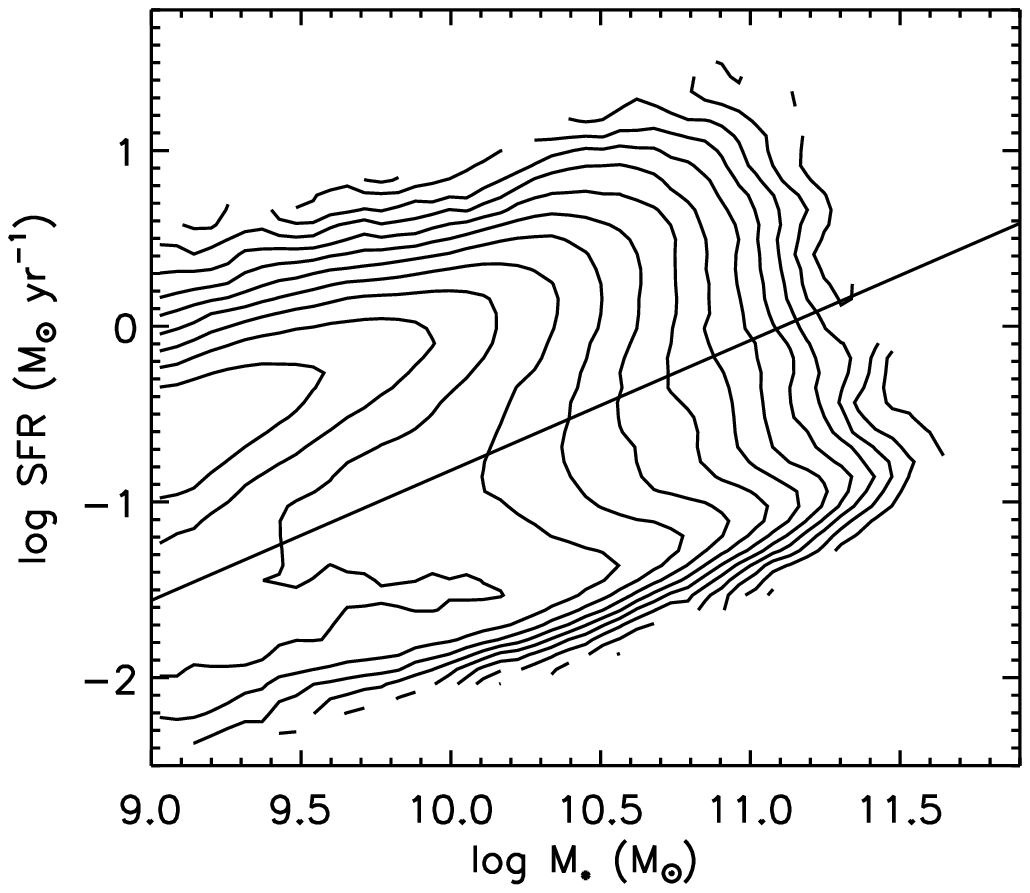}
\caption{SFR vs. $\Ms$ for our total galaxy sample.  The solid line
  marks our division between star-forming and quenched galaxies.}
\label{msfig}
\end{figure}

``Quenched'' or ``quiescent'' galaxies are those with SFR below
$\log~{\rm SFR} = 0.74\log \Ms - 8.22$.  This division, \refc{which was 
determined by eye}, is shown in
\fig{msfig}.  

\subsection{The group catalogue and halo masses}
\label{groupcat}

\cite{yang12} constructed a group catalogue and estimated group halo
masses for the SDSS DR7 sample based on the analysis of \cite{yang07},
and we use this catalogue with a slight modification for
self-consistency.  We briefly describe their group finder and
halo mass estimation below.

The group finder consists of an iterative procedure that estimates the
number density contrast of dark matter particles based on the centres,
sizes and velocity dispersion of group members, assuming a spherical
NFW profile \citep{NFW97}.  When they ran their group finder on an SDSS
mock catalogue, they successfully selected
more than $90\%$ of true haloes more massive than $10^{12}\Msun$.

With their constructed group catalogue, \cite{yang07} estimated halo
masses by rank-ordering the groups by group stellar mass.  They
assigned halo masses to the groups by rank, assuming a $\Lambda$CDM mass
function within the observed volume.  Running their group-finding and
halo abundance matching algorthim on the mock catalogue, they found an
rms scatter between real and assigned halo masses of about 0.3 dex.

The stellar masses used by \cite{yang07,yang12} are computed using the
relations given by \cite{bel03}, which can underestimate $\Ms$ for
dusty, star-forming galaxies (for a typical dust model with
attenuation of 1.6 and 1.3 in the $g$ and $r$ bands, $\Ms$ will be
underestimated by 0.2 dex; see Bell et al. 2003).  Underestimating
$\Ms$ for star-forming galaxies but not for quiescent galaxies has the
effect of exaggerating quenching trends with $\Ms$ at fixed $\Mh$.
Therefore, using the group catalogue of \cite{yang12}, we recompute
group masses using the stellar masses described in \sec{mssdss}, which
are estimated via SED fitting that incorporates dust.  
\refc{Note that this calculation does not modify the original group catalogue, 
\ie, it does not recompute group members based on the new estimates of $\Mh$
and $\Rvir$ as done in \cite{yang07}.}
\sfc{In addition to 
removing biases for specific classes of galaxies, this approach puts 
the stellar masses for the group galaxies on the same basis as used 
for the general stellar mass catalog, thus eliminating systematic 
errors between them that could eventually percolate into the halo masses.}

To account for missing members, we applied the
same correction as in \cite{yang07} (\ie, their
$g(L_{19.5},L_{\rm lim})$ factor), and use the same technique for
determining the sample groups that are complete to certain redshifts.
(Refer to their paper for details.)  Then we computed the
\cite{tinker08} mass function using the \cite{eisenstein98} transfer
function to assign halo masses to the rank-ordered group masses.  Our
estimates for halo mass are consistent with their estimates above $\Mh
= 10^{12.6}\Msun$.  In the range $12.2 < \log \Mh/\Msun < 12.6$, our
estimates are on average 0.03-0.05 dex higher than theirs, which is
expected given that dusty star-formers are found in this range.

This catalogue from \cite{yang12} contains 633 310 galaxies, of which 176 588 
make the redshift, mass and {\tt sfflag} cuts described above.  136 825 
reside in haloes with $\Mh > 10^{11.8}\Msun$.  In our analysis
we will focus our discussion on those with $\Mh > 10^{12}\Msun$ since
the comparison with the mock catalogue produced acceptable scatter
above this limit.  Below this, $\lmh$ corresponds more or less
one-to-one with the stellar mass of each galaxy since the vast
majority of these groups contain one member. 

Above $\mh < 10^{11.8} \Msun$, 85 749  
galaxies are the most massive member of their group (including groups
of only one member).  Of these, 83 391  
are also the nearest galaxy to their mass-weighted centre.  We define
these galaxies to be ``central'' galaxies.  The other most-massive
galaxies are excluded to avoid potentially unrelaxed groups (refer to
\citealp{cibinel13,carollo13b}).

This leaves 50 963 
that are not the most massive member.  Of these, a full 29 157  
reside in groups in which the most massive member is not
the nearest to its group's mass-weighted centre, \ie, in potentially
unrelaxed groups.  Examining the spatial distribution of a small
sample of these groups, we have found that sometimes a small satellite
is the nearest to the mass-weighted centre due to projection.  Since
excluding all galaxies in potentially unrelaxed groups would
drastically cut the satellite sample (and unnecessarily for some
groups), we decided to define satellites as those that are 1) in
``relaxed'' groups (those in which there is a central as defined
above) and are not the centrals themselves, or 2) ranked third massive
in the group or lower if they are in potentially unrelaxed groups.
These criteria yield a sample of 48 598 
satellites.  The idea behind
the second criterion is that the group halo is most likely associated
with the two most massive members.  However if we restricted our
sample only to the first criterion, our results, though much noisier
\sfc{(after applying the compactness cuts below)}, would remain
qualitatively unchanged.

We define the relative group-centric distance of satellites as the ratio of
the projected distance $\dproj$ of each satellite to the mass-weighted
group centre and the virial radius $\Rvir = 120
(\Mh/10^{11}\Msun)^{1/3} {\rm kpc}$ (\eg, \citealp{dekbir06}).

\subsection{Galaxy Compactness}
\label{morph}

In order to probe quenching mechanisms which result in centrally
concentrated galaxies, we compute the surface density within the inner
kiloparsec $\sigone$ in the following way.  We retrieved surface
brightness profiles in the $ugriz$ bandpasses from the SDSS DR7
Catalog Archive Server and corrected them for galactic extinction
using the extinction tags in the {\tt SpecPhoto} table.  Then we used
the flux within each radial bin and in each band as input into the
K-correction utilities of \cite{blanton07} (v4\_2) to compute the
stellar mass profile.  We then summed the bins to compute the
cumulative stellar mass profile and interpolated between the radial
bins to estimate the total mass within 1 kpc and compute the density
in this radius.  

Out of 897 582 galaxies with profiles retrieved from CasJobs with $z <
0.3$, 133 192 
make the cuts described above: 81 063 
centrals and 47 524 
satellites.  None of these had their largest radial bin smaller
than 1 kpc, or their smallest radial bin larger than 1 kpc.

\refc{This estimate of compactness does not account for contamination  
from near neighbours.  This contamination will be strongest for satellites of
large and dense clusters.  We checked the importance of this effect on the 
$\sigone$-sSFR relation (using an analysis similar to \sec{seeing}).  We find 
no noticeable 
difference in the mean trends between galaxies whose nearest photometric 
neighbour is less than $10''$ and those more isolated.  
Therefore, we do not attempt to correct for this 
effect.}

\refc{This estimate also does not correct for inclination, but we have  
confirmed that applying a selection on inclination ($b/a > 0.5$) does not 
change our results. }

Of potential greater importance is that this computation does not correct for
atmospheric seeing.  An estimate of the PSF for each galaxy is the
effective width of the best-fit double-Gaussian PSF model in the
centre of each SDSS frame (the {\tt psfWidth} tag in the {\tt Field}
table) \citep{fan01,stoughton02}.  Of the 81 063 
centrals in our sample, only 7572 
are on frames with PSF width less than 1 kpc.  These
are almost all below $z \sim 0.04$ and include too few 
high-mass haloes to observe quenching trends with $\Mh$.  Therefore,
for the central population, we have selected those galaxies whose PSF
widths are less than 2 kpc (57 113  
centrals).  We discuss the effects
of the PSF on our results in \sec{seeing} and attempt to correct for
it.

On the other hand, the satellite population, if limited to those whose
PSF widths are less than 1 kpc (8826) 
will contain objects in massive
haloes in sufficient numbers.  Thus our analysis of satellites will
only include these objects so that our $\sigone$ values for these
satellites will not be significantly affected by seeing.

Therefore our final sample consists of 57 113 centrals and 8826
satellites.

\section{Centrals}
\label{centralssection}

\begin{figure*}
\epsscale{2.1}
\plottwo{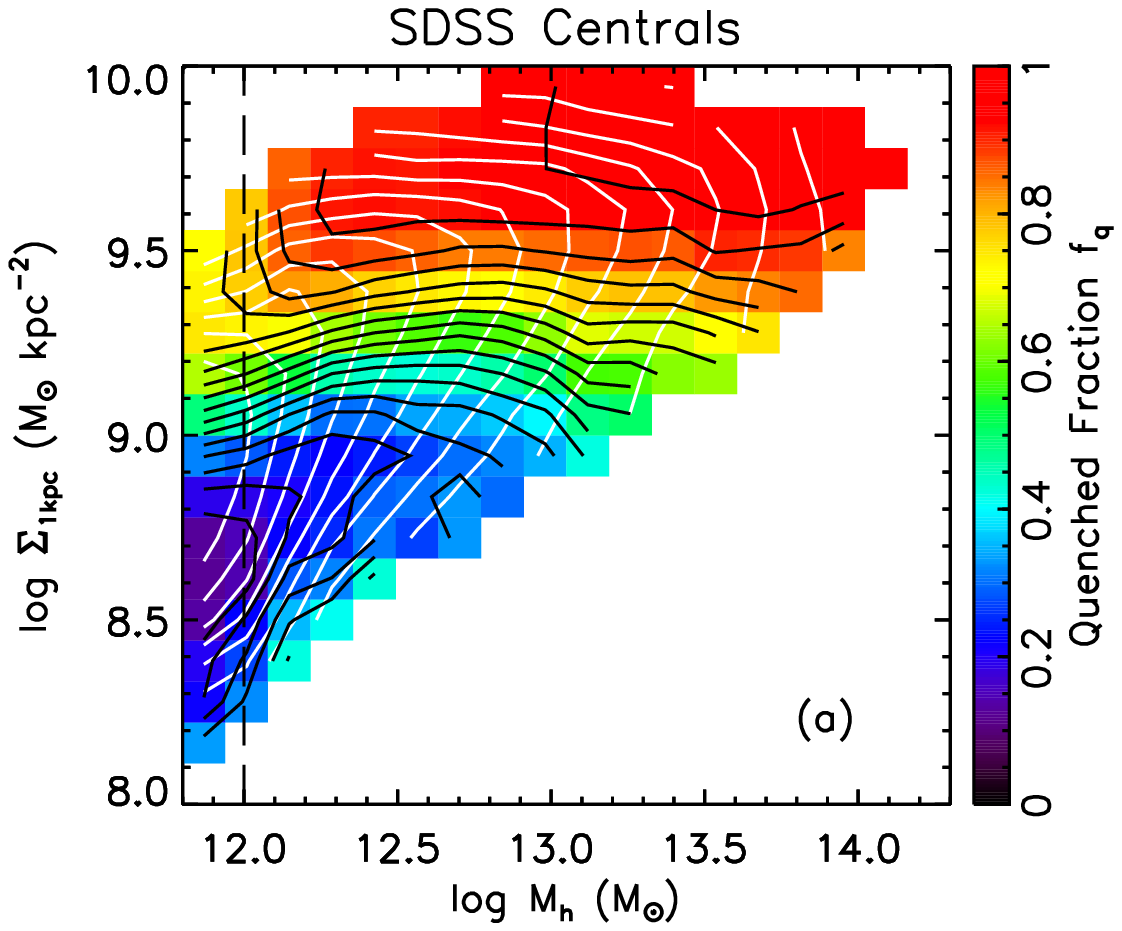}{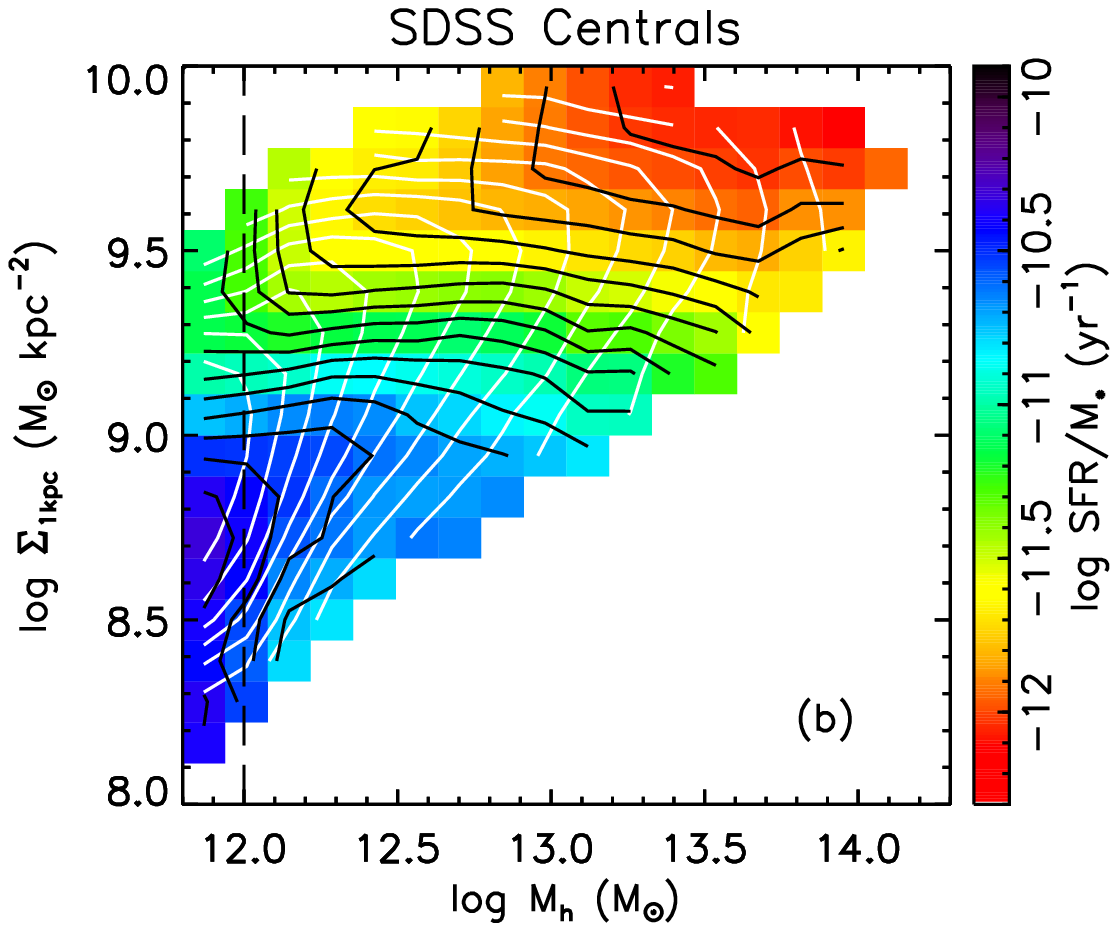}
\caption{The quenched fraction ($a$) and the mean log SFR/$\Ms$
  ($b$) in the $\sigone$-$\Mh$ plane for central
  galaxies after $\sigone$ has been corrected for PSF effects as
  described in \sec{seeing}.  The white contours represent the number
  density of galaxies per pixel with a maximum of 0.001 Mpc$^{-3}$ and
  a separation of 0.25 dex.
  The black contours follow the colour scale and are
  separated by 0.05 for $fq$ and 0.12 dex in yr$^{-1}$ for sSFR.  The
  vertical dashed line marks $\Mh = 10^{12}~\Msun$ below which the
  errors in the $\Mh$ estimates increase dramatically, and it also
  refers to the $\Mcrit$ near which halo quenching is predicted to
  become important.  The quenched
  fraction ans sSFR for centrals strongly depends on $\sigone$ at fixed
  $\Mh$ for mid-range values of $\sigone$ ($\sim 10^{9-9.4}\Msun
  \kpc^{-2}$).  
  For higher and lower values of $\sigone$, there is a gradual decrease
  in sSFR (and increase in $\fq$) up to $\Mh \sim 13$.
  (The same plots without the PSF correction are very
  similar.)}
\label{morphmass}
\end{figure*}

Here we compare the strength of the quenching correlation with galaxy
compactness to the strength of the quenching correlation with halo mass
for central galaxies.  \fig{morphmass}$a$ shows the
quenched fraction \sfc{(represented by the colour scale)} a function of the 
$\sigone$-$\Mh$ plane for
central galaxies.  ($\sigone$ is corrected for atmospheric seeing
according to \sec{seeing}, but the uncorrected plots are almost
indistinguishable.)  The white contours show the number density of
galaxies per pixel and are separated by 0.25 dex in Mpc$^{-3}$.
The black contours follow the colour scale and are separated by
0.05.  The vertical dashed line marks $\Mh = 10^{12} ~\Msun$ below
which the errors in the $\Mh$ estimates increase dramatically (this
also happens to be $\sim \Mcrit$).  \fig{morphmass}$b$
is the same as panel $a$ except the colour scale represents
the mean sSFR, and the black contours are separated by 0.12 dex in
sSFR.

\sfc{This figure shows that for central galaxies with 
$\sigone \sim 10^{9-9.4}\Msun \kpc^{-2}$, 
$\fq$ and sSFR are strongly correlated with $\sigone$ with only weak
$\Mh$ dependence.  
In particular, the range $\sigone \sim 10^{9-9.4}\Msun \pkpc2$ seems to 
divide the galaxies into those with  
with high sSFR below this range and low sSFR above this range.
However, at higher and lower ranges of $\sigone$, 
sSFR and $\fq$ (the latter especially around $\Mcrit$)
are strong functions of $\Mh$, and mostly independent of $\sigone$.
These roughly vertical and horizontal contours of quenching in this plane
seem to be evidence for
two modes of quenching for centrals, one related to the halo and one
related to galaxy compactness.}

\begin{figure*}
\epsscale{2.1}
\plotthree{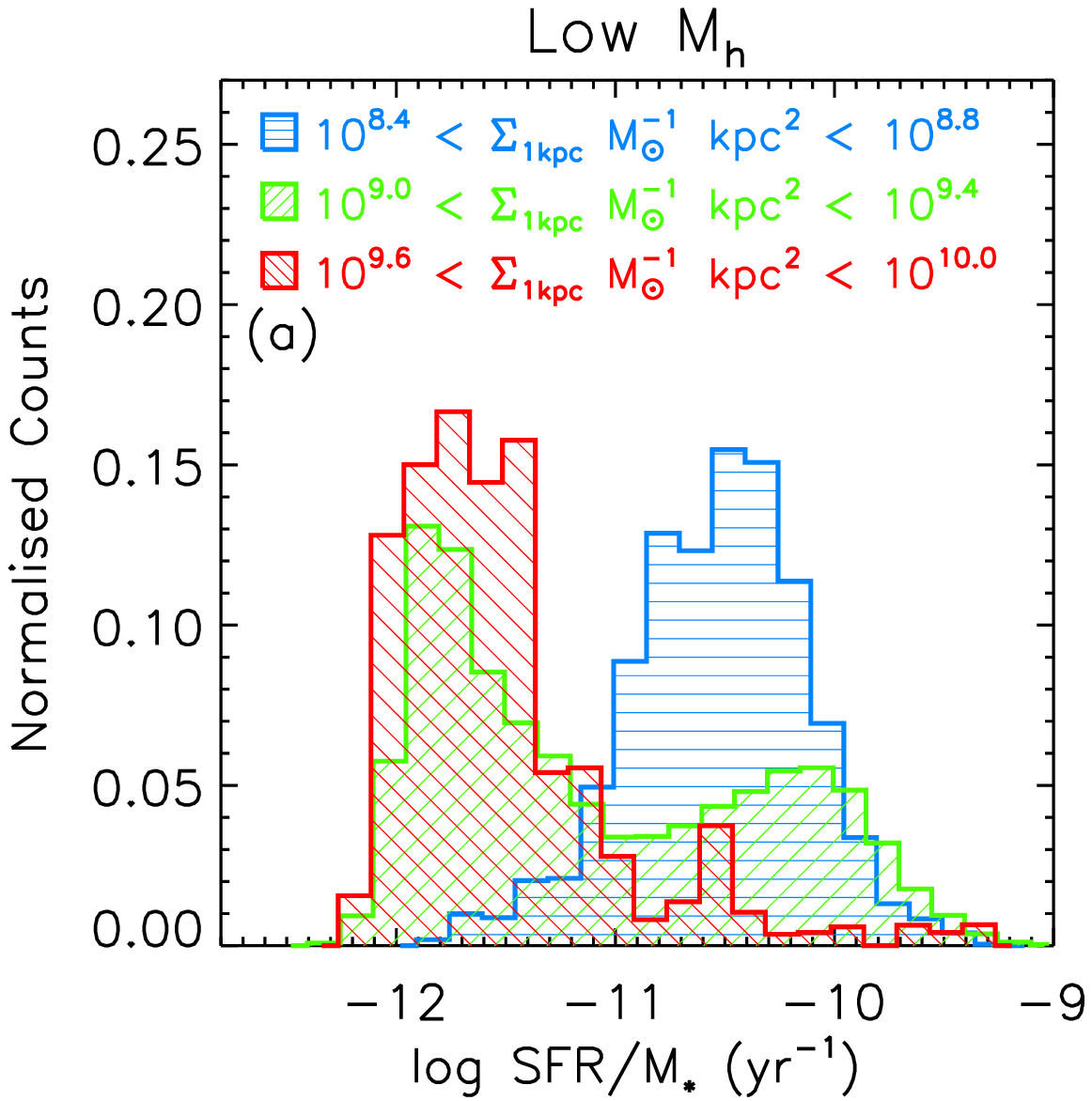}{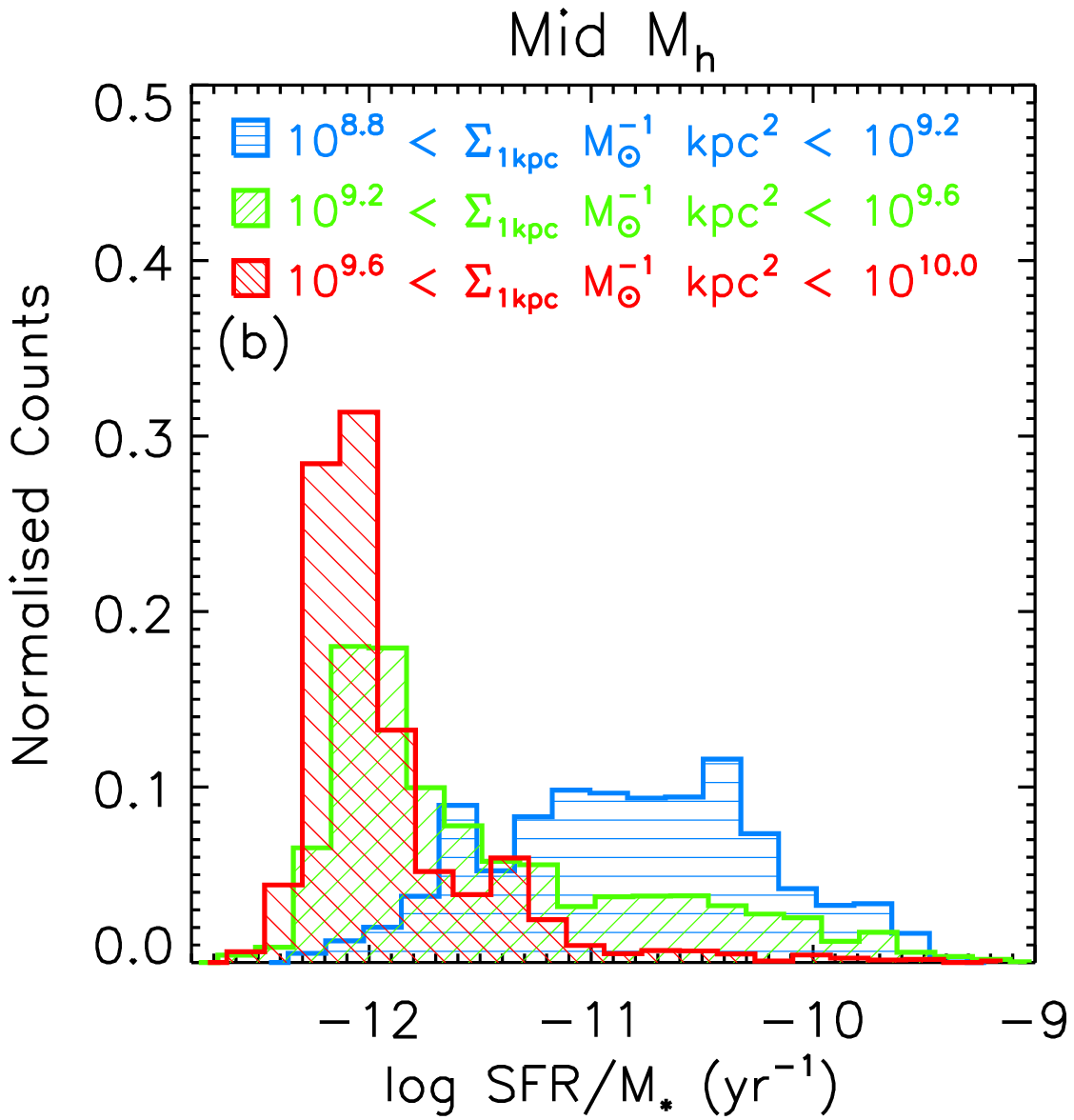}{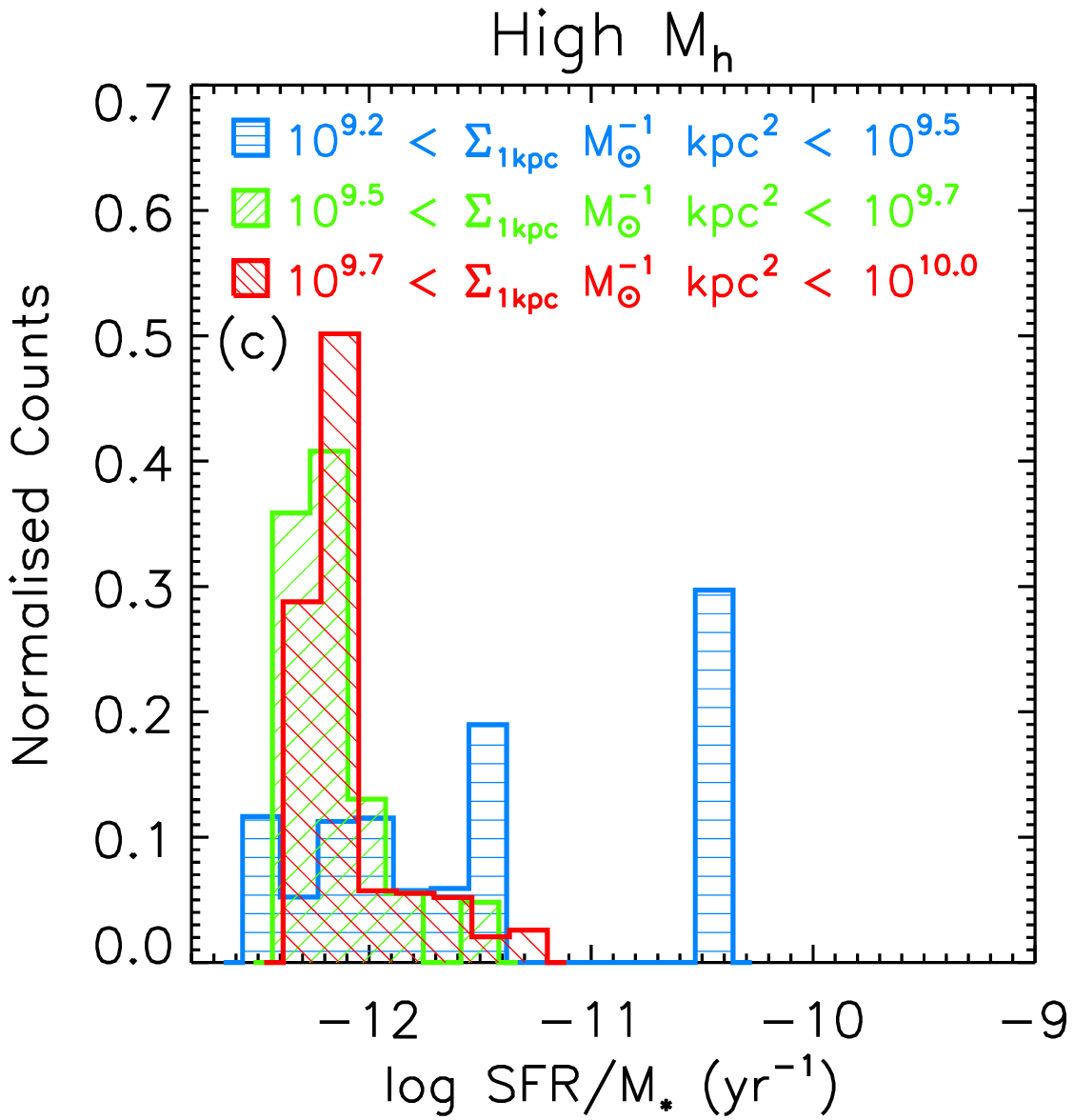}
\caption{\refc{Distribution of sSFR for central galaxies in different bins
  of $\sigone$ (represented by different coloured/hashed histograms) for low 
(panel $a$: $12.0 < \log \Mh/\Msun < 12.3$), 
  mid-range (panel $b$: $12.7 < \log \Mh/\Msun < 13$, and high (panel $c$: 
$13.7 < \log \Mh/\Msun < 14$) values of $\Mh$. 
  Varying $\sigone$ at fixed $\Mh$ changes the relative
  frequencies of galaxies with high- and low-sSFR.}}
\label{lssfrhists_fixedmh}
\end{figure*}

\begin{figure*}
\plotthree{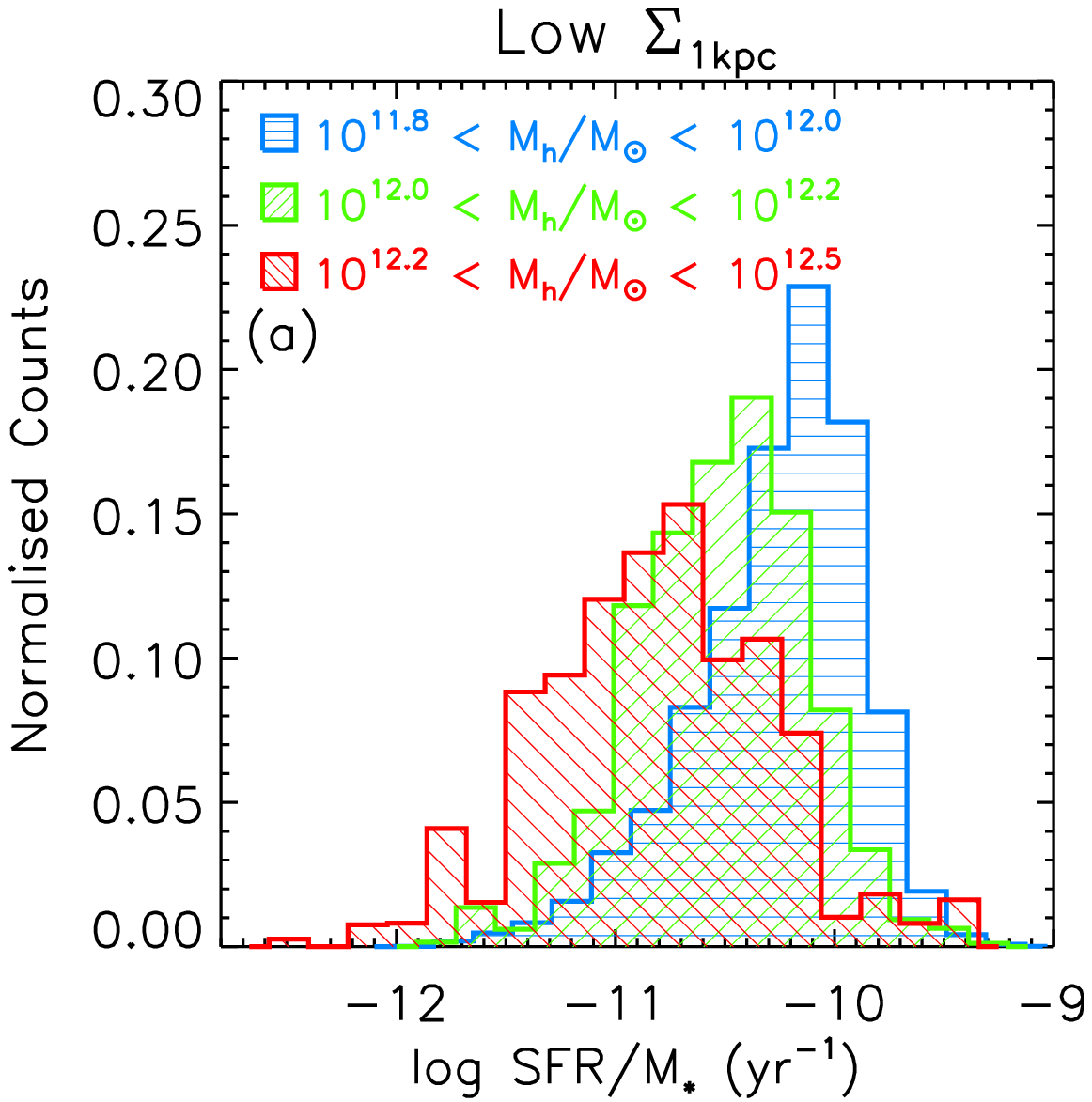}{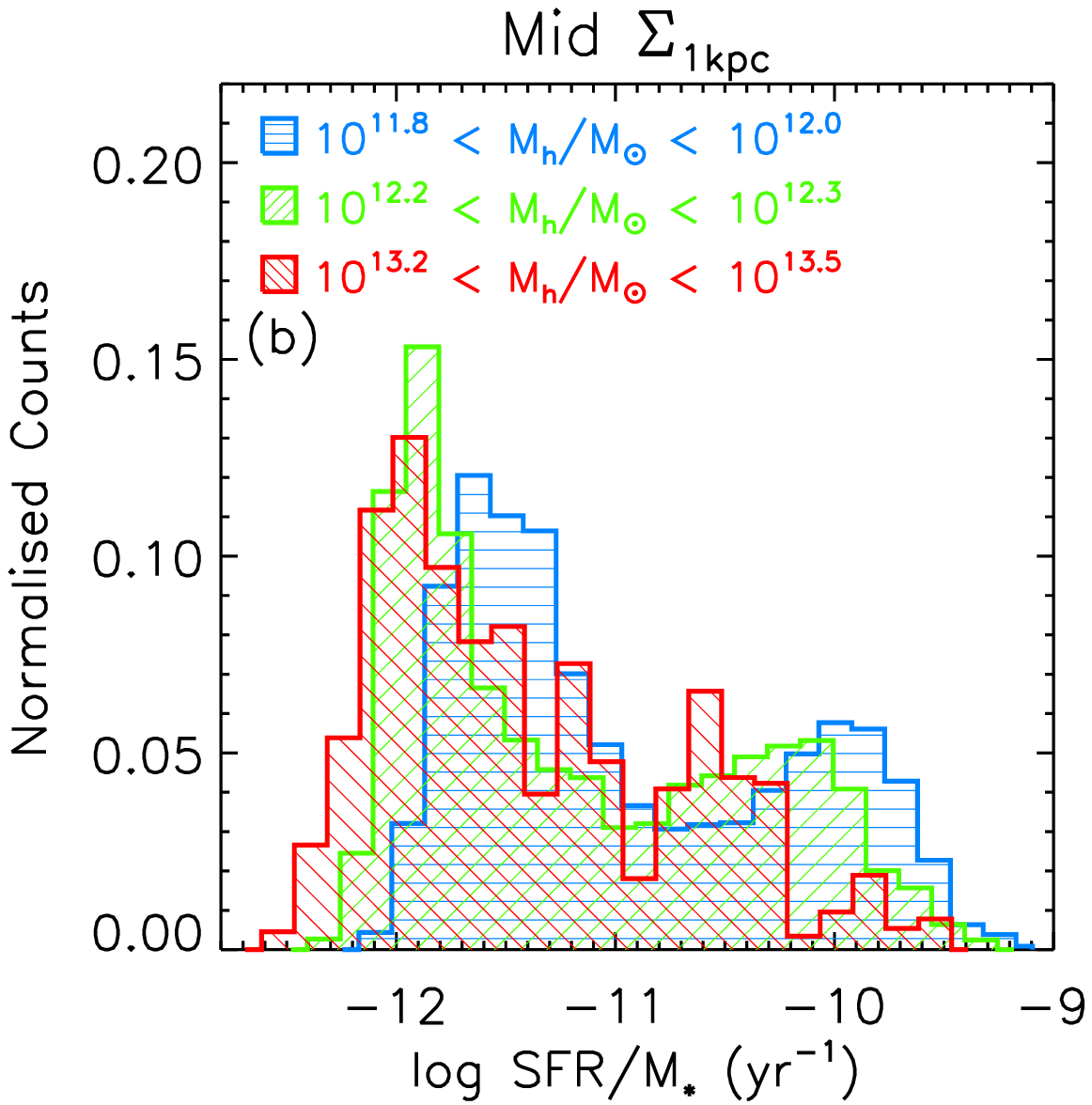}{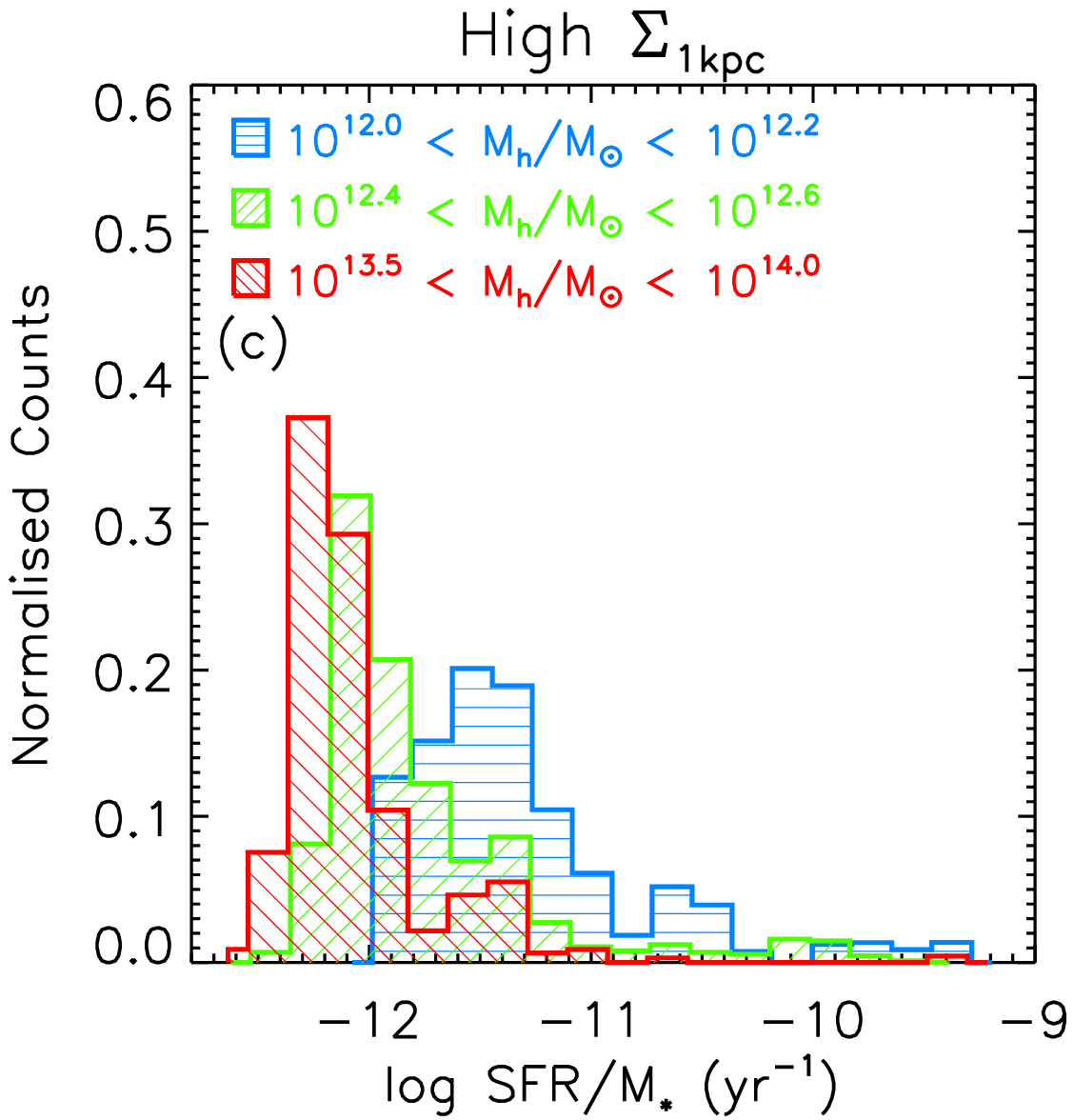}
\caption{\refc{Distribution of sSFR for central galaxies in different bins
  of $\Mh$ (represented by different coloured/hashed histograms) for 
  low (panel $a$: $8.3 < \log \sigone/\Msun \pkpc2 < 8.8$), 
  mid-range (panel $b$: $9.0 < \log \sigone/\Msun \pkpc2 < 9.5$),
  and high (panel $c$: $9.6 < \log \sigone/\Msun \pkpc2 < 10$) values of 
$\sigone$.  Varying $\Mh$ at fixed $\sigone$
  shifts the whole distribution of sSFR to lower values without a significant 
change in its 
shape.}}
\label{lssfrhists_fixedsigma}
\end{figure*}

\begin{figure*}
\begin{center}
\epsscale{1.8}
\plottwo{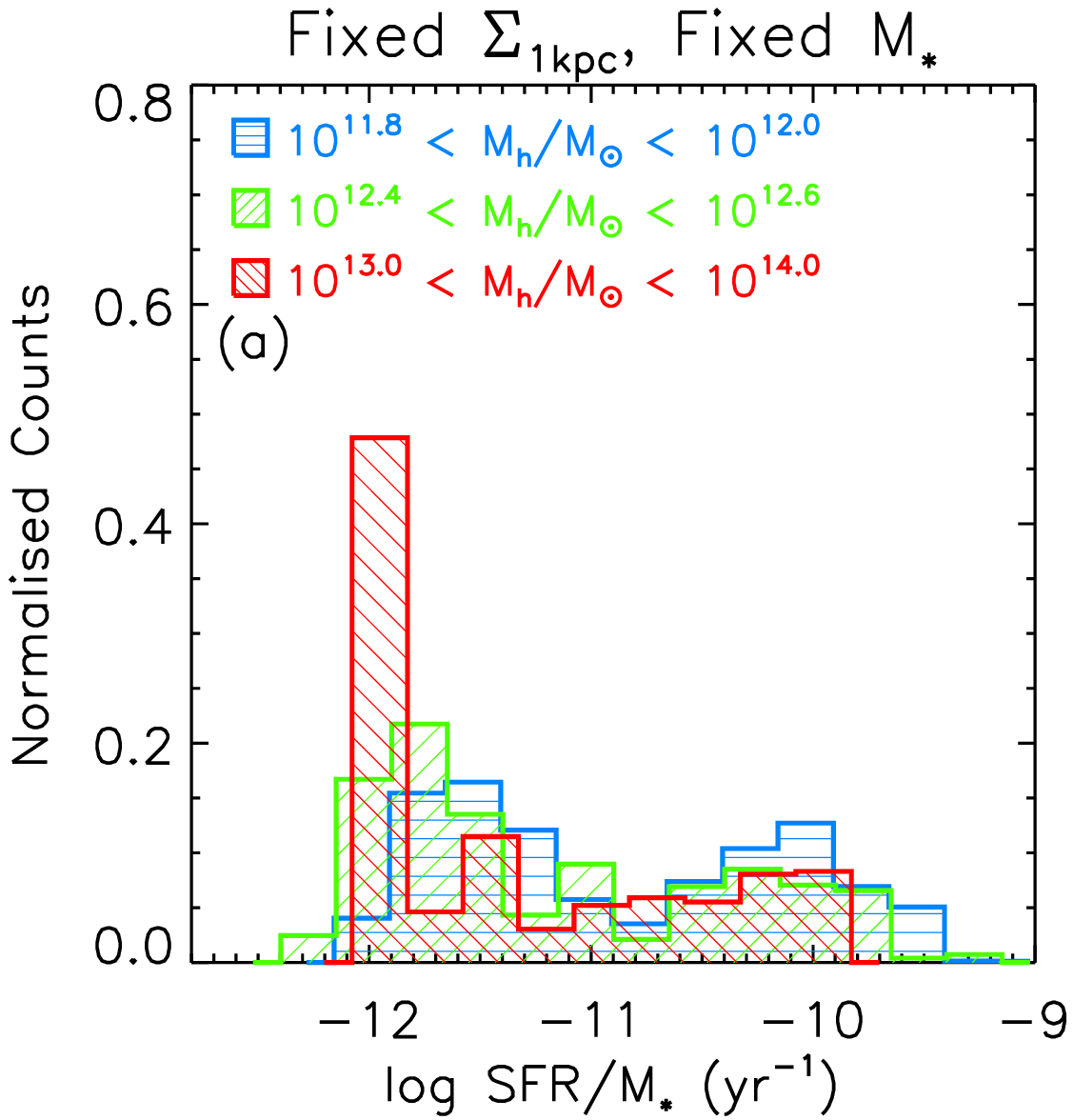}{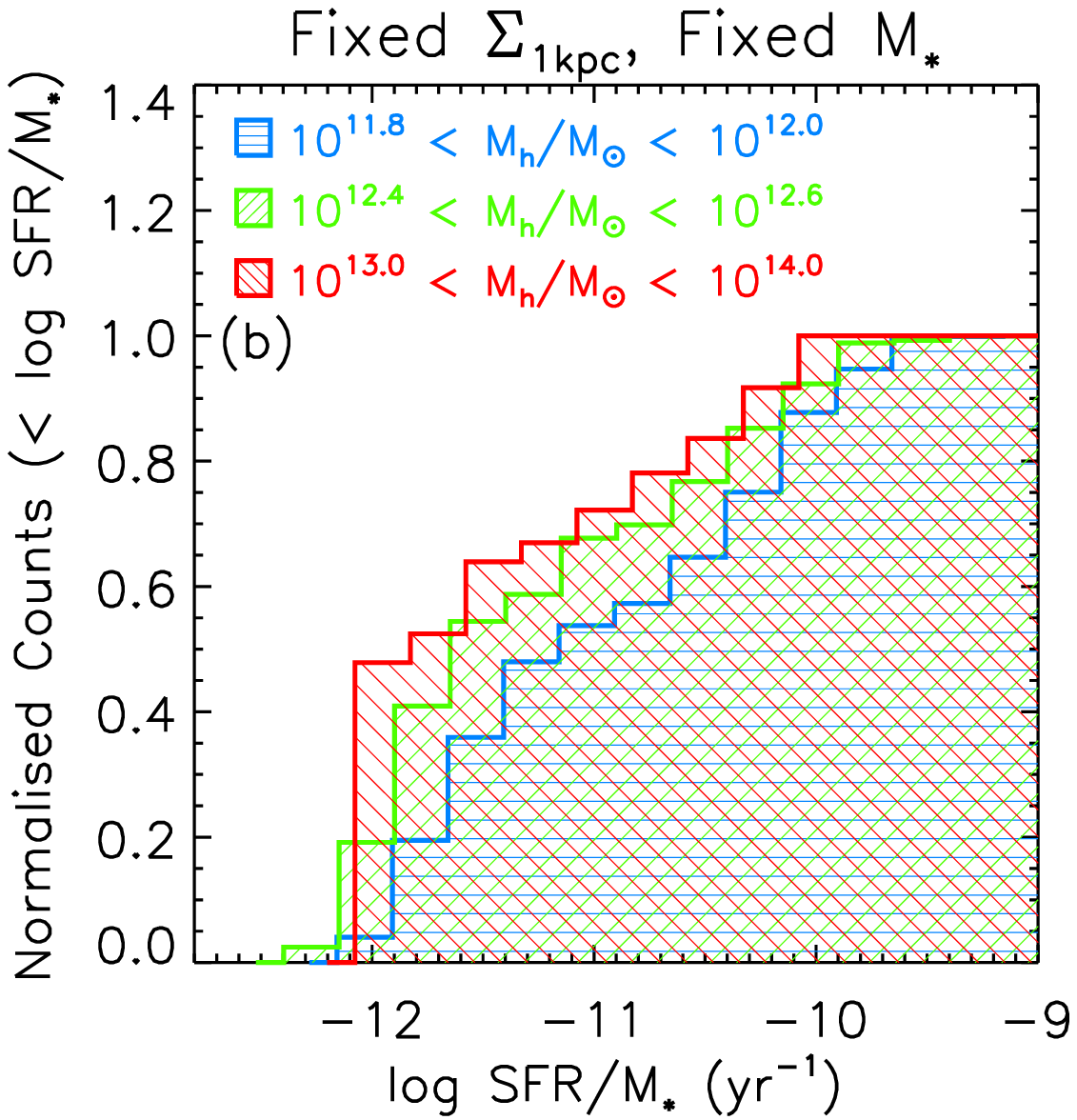}
\caption{The distribution of sSFR ($a$) and the cumulative
  distribution of sSFR ($b$) for central galaxies at fixed
  $\sigone$ \refc{($9 <  \log \sigone/\Msun \pkpc2$)} and fixed $\Ms$ 
\refc{($10.3 < \log \Ms/\Msun < 10.6$)} for
  different bins of $\Mh$, represented by the different coloured/hashed histograms.  
  Increasing $\Mh$ seems to shift the sSFR
  distribution to lower values.}
\label{lssfrhists_fixedMs}
\end{center}
\end{figure*}

When considering \fig{morphmass}, keep in mind that sSFR is a bimodal
quantity and the panels are measuring different aspects of this
bimodality.  Thus we complement these figures with the entire
distribution of sSFR in \twofigs{lssfrhists_fixedmh}{lssfrhists_fixedsigma}.  
In 
\fig{lssfrhists_fixedmh}, we show the sSFR distribution for different bins of
$\sigone$, \refc{represented by the different coloured/hashed histograms, 
in three panels of fixed $\Mh$, which are three vertical slices} 
of \fig{morphmass}$b$.
This figure shows that varying $\sigone$ changes the shape of
the sSFR bimodality strongly in that higher $\sigone$ results in more
galaxies in the quenched peak.  This is seen in \fig{morphmass}$a$
as a strong increase in $\fq$.  This also explains the strong decrease
in mean sSFR with $\sigone$ for mid-range values of $\sigone$, \ie,
the green shaded region in \fig{morphmass}$b$.  This region does
not represent a peak of galaxies in the ``green valley'' of the sSFR
bimodality, but rather a region where there are comparable numbers of
galaxies on either side of the bimodality.

In \fig{lssfrhists_fixedsigma} we show the sSFR distribution
for different bins of $\Mh$, \refc{represented by different coloured/hashed histograms,
in three panels of fixed $\sigone$, which are three horizontal
slices of \fig{morphmass}$b$}.  In contrast to $\sigone$, this plot
shows that varying $\Mh$ does not strongly change the shape of the
bimodality, but rather shifts the whole distribution of sSFR downward
with increasing $\Mh$ (at fixed $\sigone$).  This is seen in
\fig{morphmass}$b$ as the decrease in sSFR for low and high
$\sigone$.
In fact, the highest bin of $\Mh$
for the low-$\sigone$ galaxies (the red histogram in \fig{lssfrhists_fixedsigma}$a$)
is not bimodal in sSFR, but rather
straddles the ``green valley''.  In other words, when quenching by
$\sigone$ is not present, neither is the bimodality.  Halo-related
quenching, in the absence of $\sigone$-quenching, reduces sSFR more
continuously (rather than abruptly shifting the galaxy from
star-forming to quenched).

It may be noted that the low-sSFR peak in the sSFR bimodality
represents more of an upper limit instead of actual measurements of
sSFR.  The real sSFR values may in fact decrease with increasing
$\sigone$ even though we do not see this in \fig{lssfrhists_fixedmh}.
However these upper limits {\it are} seen to decrease with increasing
$\Mh$, showing that $\Mh$ more strongly predicts sSFR than $\sigone$
for quenched galaxies.  At the very least, one can conclude from
\twofigs{lssfrhists_fixedmh}{lssfrhists_fixedsigma}
(and \fig{morphmass}) that $\sigone$ predicts the
shape of the distribution of sSFR (\ie, the number of galaxies on
either side of the bimodility) whereas $\Mh$ predicts the position in
sSFR of the entire distribution.

It may also be noted that the different bins of $\Mh$ in 
\fig{lssfrhists_fixedsigma} also represent different bins of $\Ms$ since
for central galaxies, these are closely correlated.  (The fixed bins of
$\Mh$ in each panel of \fig{lssfrhists_fixedmh} do represent roughly fixed bins
of $\Ms$.)  Thus a shift of the sSFR distribution shows that
the downward slope of the sSFR-$\Ms$ relation is still negative after
selecting galaxies with a fixed $\sigone$.  However, we show in
\fig{lssfrhists_fixedMs} that $\Mh$ also reduces sSFR independently of
$\Ms$.  Panel $a$ shows the distribution of sSFR for the same
range of $\sigone$ as in \fig{lssfrhists_fixedsigma}$b$, but also for fixed
$\Ms$ ($10^{10.3-10.6} \Msun$).  Since we have greatly
reduced the sample, the distributions are rather noisy, so we also
show the cumulative distribution of sSFR in 
\fig{lssfrhists_fixedMs}$b$.  These panels show that even at fixed $\Ms$ and 
fixed
$\sigone$, increasing halo mass seems to shift the sSFR distribution
to lower values.  \refc{The same shift is seen when restricting the $\Ms$ 
ranges of \fig{lssfrhists_fixedsigma}$a$ and $c$.
A two-sided Kolmogorov-Smirnov test shows that the 
probability that the red and blue histograms in \fig{lssfrhists_fixedMs} are 
drawn from the same distribution is less than $10^{-5}$.  We also performed the 
same analysis on the distribution of $\Delta$sSFR, \ie, the distance of the 
sSFR from the 
dividing line in \fig{msfig} and find the similar results.  
} 

\begin{figure}
\begin{center}
\includegraphics[width=0.47\textwidth]{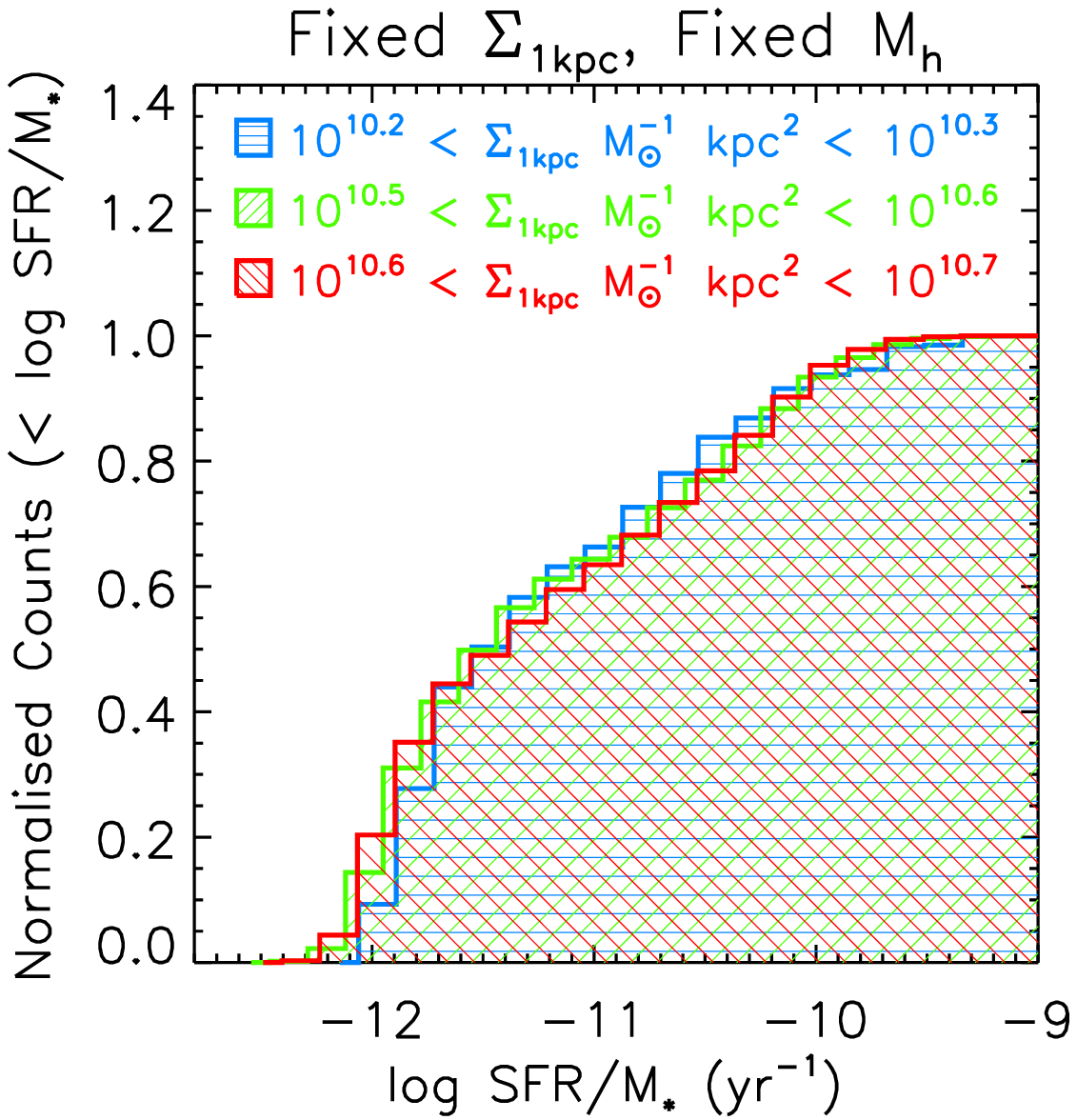}
\caption{\arxc{The cumulative sSFR distribution in three bins of $\Ms$ after 
fixing 
$\sigone$ ($10^{9-9.5} \Msun \pkpc2$) and fixing $\Mh$ ($10^{12.0-12.3} 
\Msun$). 
$\Ms$ does not significantly alter the sSFR distribution.}}
\label{fixedsigmafixedmh}
\end{center}
\end{figure}

\refc{To further explore the role of $\Ms$ in determining 
sSFR, we show in \fig{fixedsigmafixedmh} the cumulative distribution in 
three bins of $\Ms$ in a slice of both $\sigone$ ($10^{9-9.5} \Msun \pkpc2$) 
and 
$\Mh$ ($10^{12.0-12.3} \Msun$).  
The three distributions lie on top of each other.  
A two-sided Kolmogorov-Smirnov test shows that the 
probability that the red and blue histograms are 
drawn from the same distribution is as high as 0.13.  
However, note that fixing the range of $\Mh$ 
leaves only a small range of $\Ms$ in contrast to the other way around 
(\fig{lssfrhists_fixedMs}) due to 
the shape of the $\Ms$-$\Mh$ relation \citep{yang07,moster10,woo13}.
Thus $\Ms$ seems to provide no additional information for quenching to the 
combination of $\sigone$ and 
$\Mh$, whereas $\Mh$ does add information over $\Ms$ (\fig{lssfrhists_fixedMs}), 
in agreement with \cite{woo13}.
} 

How important is halo quenching?  About 36\% of diffuse galaxies with
$12.2 < \log \Mh/\Msun < 12.5$ are quenched
(red histogram of \fig{lssfrhists_fixedsigma}$a$).  Assuming that $\sigone$-related
quenching and halo quenching are the only quenching mechanisms, these
are quenched by the halo alone.  Furthermore, of all the quenched
galaxies in \fig{morphmass} (namely $\Mh > 10^{11.8} \Msun$), 21\% lie
below $\sigone = 10^9 \Msun \kpc^{-2}$ and these must be quenched by
the halo alone.  (Since the threshold mass for halo quenching can vary
over a wide range around $10^{12} \Msun$, some of the haloes of $\sim 10^{11.8}
\Msun$ can easily be quenching via the halo.)

Similarly, those that are quenched by $\sigone$-related processes
alone would be those with mid-to-high $\sigone$ and low $\Mh$, such as
the left peak of the blue histograms in \fig{lssfrhists_fixedsigma}$b$ and $c$.
Of all quenched galaxies in
\fig{morphmass}, 35\% are above $\sigone = 10^9 \Msun \kpc^{-2}$ and
below $\Mh =10^{12}\Msun$.  These can be suspected of being quenched
by $\sigone$-related quenching only, not yet being in massive enough haloes
for maintaining halo quenching.  However $\Mh$ for these galaxies is
high enough for the halo to possibly also play a role in their
quenching.

Most of the quenched centrals have both high $\sigone$ and high $\Mh$
(79\% with $\sigone > 10^9 \Msun \kpc^{-2}$ and $\Mh >
10^{11.8}\Msun$).  These seem to have experienced both quenching
mechanisms, though one may have occurred before the other.

These results taken together point to the presence of both
compactness-related quenching and halo quenching, and to the 
nature of their quenching roles.  If the
increase of $\fq$ measures the transfer of galaxies from one side of
the galaxy bimodality to the other, while the decrease of sSFR relates
to the fading of star formation, then for central galaxies, the
process related to 
compactness plays the transferring role, while the halo plays the fading
role.  This will be discussed further in \sec{discussion}.

\subsection{The Effect of Atmospheric Seeing}
\label{seeing}

Since we have not limited the sample of centrals to those where the
PSF width is less than 1 kpc (but rather to those for which the PSF
width is less than 2 kpc), it is important to discuss potential
effects that atmospheric seeing may have on our results.  Since the
PSF tends to move light outward in monotonically decreasing surface
brightness profiles, uncorrected profiles will underestimate
$\sigone$.  Furthermore, since the PSF width is smaller in red
bandpasses than in blue bandpasses, \jwc{this could potentially have
one or both of two opposite consequences.  First,}
the uncorrected $\sigone$ for blue
galaxies will be underestimated to a greater degree than for red
galaxies.  In other words, if colour roughly translates to sSFR, the
effect of seeing is to artificially raise the mean sSFR in low bins of
$\sigone$.  Similarly, if colour roughly translates to the quenched
fraction, the effect of seeing is to artificially steepen the gradient
of $\fq$ with $\sigone$.  \jwc{Second, in a given galaxy, 
bluer light moves outward more than red light, leaving its centre 
appearing redder, and thus appearing to have higher mass-to-light ratio.  
Thus, the PSF can artificially raise $\sigone$}.

\begin{figure}
\begin{center}
\includegraphics[width=0.47\textwidth]{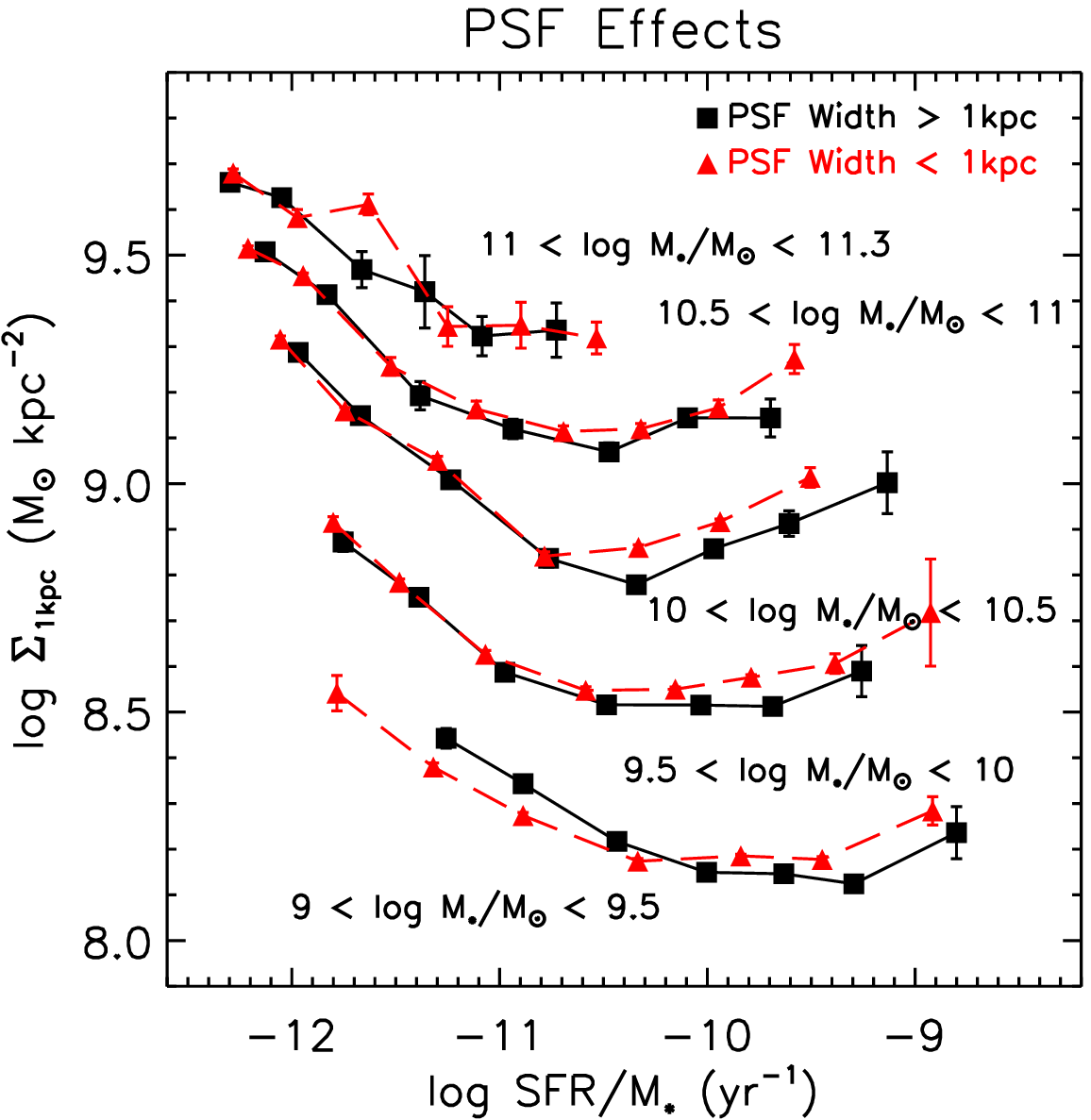}
\caption{Measured values of the mean $\sigone$ as a function of sSFR
  and $\Ms$ comparing those galaxies whose PSF widths are larger than
  1 kpc (black squares) to those galaxies whose PSF widths are smaller
  than 1kpc (red triangles).  The error bars are errors on the mean.
  The effect of seeing is to underestimate the measured $\sigone$,
  especially for low-mass star-forming (blue) galaxies.  The large-PSF
  group of low-mass passive galaxies have higher $\sigone$ than the
  small-PSF group, likely because the former includes galaxies at
  higher $z$ where low surface- brightness galaxies drop from the
  sample.}
\label{psfeffects}
\end{center}
\end{figure}

How important are these effects?  One way to measure this is to look at
the mean $\sigone$ as a function of sSFR, comparing galaxies with good
PSF widths ($< 1$kpc) to galaxies with large PSF widths ($> 1$kpc).
We show this comparison in \fig{psfeffects} in several bins of $\Ms$.
The red triangles are those with good PSF widths, and the black
squares are those with large PSF widths.  This figure shows that
galaxies with high sSFR (\ie, those that are bluer) have higher
measured $\sigone$ when the PSF widths are small compared to those
whose PSF widths are large.  In other words, the PSF causes an
underestimate of $\sigone$ for bluer galaxies as expected.  The
difference between the mean $\sigone$ is at the most about 0.1 dex and
is larger for less massive galaxies.  For the lowest mass bin,
\fig{psfeffects} also shows that galaxies with low sSFR have higher
measured $\sigone$ when the PSF is large compared to the those whose
PSF is small (\ie, the black line is higher than the red line).  
\jwc{These may either be because of the second effect of the PSF described 
above, or 
because the large-PSF group on average higher $z$ (within the same $z$-bin)} where
low-surface-brightness galaxies drop from the sample.

Thus, to create \fig{morphmass}, we estimated a correction to
$\sigone$ based on \fig{psfeffects}.  For each bin of log $\Ms$, we
calculated the difference between the red and black lines in
\fig{psfeffects}.  This difference is an increasing function of log
sSFR which is roughly linear.  A linear least squares fit to the
differences resulted in an approximate upward correction to $\sigone$
as a function of sSFR in bins of $\Ms$.  We applied this correction to
those galaxies whose PSF widths are larger than 1 kpc.  

After applying these rough corrections, we compared \fig{morphmass} to
the same plot without corrections (not shown here) and find that the
differences are almost imperceptible.  Thus the PSF seems to have a
small effect on the broad trends of $\fq$ and sSFR for most of the
$\sigone$-$\Mh$ plane if we select PSF widths $<$ 2 kpc (which is our
sample of centrals).  The most significant change is that the mean
$\fq$ is slightly lowered (by less than 0.05) at high $\sigone$ below
$\Mh \sim 10^{12}\Msun$.  We expect the PSF effects to become more
significant for selections that include larger PSF widths and higher
$z$.

\section{Satellites}
\label{satellites}

\cite{woo13} showed that the quenched fraction for satellites depends
strongly on group-centric distance $\Dist$ and $\Mh$.  These authors
showed that $\fq$ also depends on $\Ms$, but only in the outer regions
of groups.
Position in the host halo correlates with
time after infall so this result suggests that satellites behave as
centrals (ignoring the host halo) for some time after they fall in
\citep{wetzel12}.

62\% of satellites reside in the outer regions ($\log (\Dist) > -0.5$) while
only 8\% of them live in the inner regions ($\log (\Dist) < -1$).  Therefore
quenching studies which do not separate satellites into regional bins
(or even worse, combine all centrals and satellites) will miss the
strong $\Mh$ signal seen in \cite{woo13}.  Therefore, when comparing the effects
of the halo and galaxy compactness on quenching, we must keep these
regional differences in mind.

\begin{figure*}
\includegraphics[width=\textwidth]{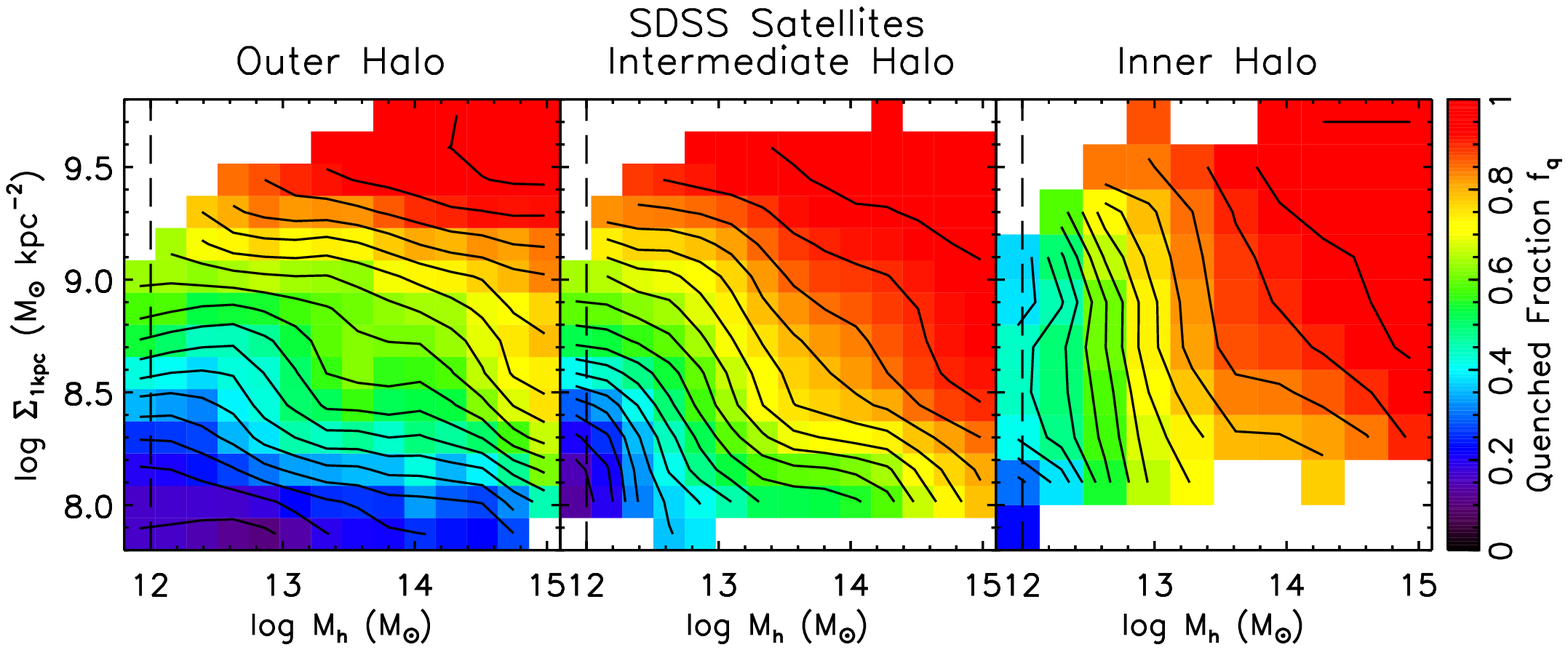}
\includegraphics[width=\textwidth]{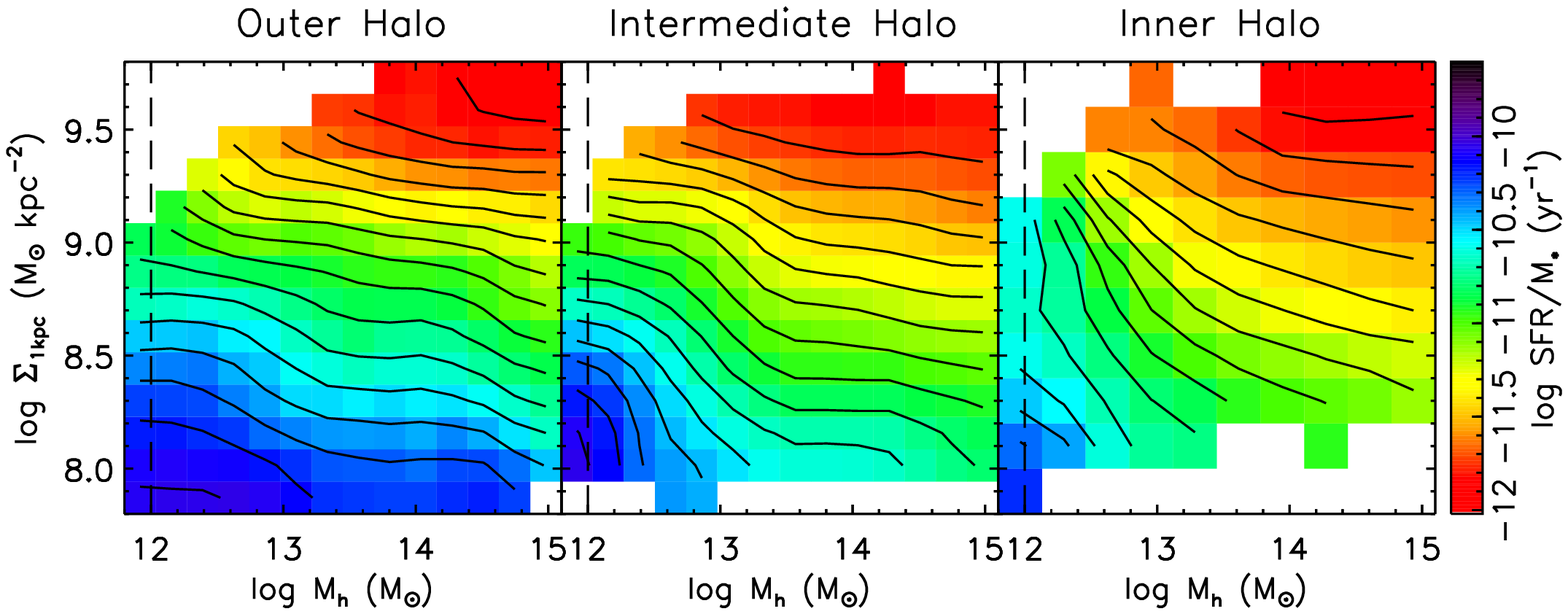}
\caption{The quenched fraction (top) and sSFR (bottom) of satellites
  as a function of the $\sigone$-$\Mh$ plane in three bins of
  group-centric distance.  The ``Outer Halo'' refers to $\log (\Dist)
  > -0.5$, ``Intermediate Halo'' refers to $-1.0 < \log (\Dist) <
  -0.5$ and the ``Inner Halo'' refers to $\log (\Dist) < -1.0$.  This
  sample is limited to only those satellites for which 1 kpc is
  greater than the PSF width.  The white contours represent the number
  density of galaxies per pixel and are separated by 0.25 dex in
  Mpc$^{-3}$ with maxima of $8\times10^{-4}$,
  $4\times10^{-4}$, and $8\times10^{-5}$ Mpc$^{-3}$ in each
  panel from left to right. 
  The black contours follow the colour scale and are 0.05
  apart for $\fq$ and 0.12 dex in yr$^{-1}$ for sSFR.  The vertical
  dashed line marks $\Mh = 10^{12}~\Msun$ below which the errors in
  the $\Mh$ estimates increase dramatically.  $\fq$ depends
  predominantly on $\Mh$ in the inner regions of haloes, while sSFR
  depends on both $\Mh$ and $\sigone$ in this region.}
\label{morphmhdbinsgoodpsf}
\end{figure*}

\fig{morphmhdbinsgoodpsf} (top) shows the quenched fraction of
satellites as a function of the $\sigone$-$\Mh$ plane in three bins of
$\Dist$.  This figure shows that $\fq$ strongly depends on $\sigone$
for satellites in the outer regions of groups, just as it does for
centrals.  However even here in the outer halo, the influence of the
host halo is non-negligible \sfc{since $\fq$ increases slightly with
  $\Mh$ at fixed $\sigone$}.  The $\Mh$ dependence of the quenched
fraction becomes much more dominant in the inner regions of haloes
where the $\sigone$ dependence of $\fq$ almost disappears.

Note also that \sfc{the quenched fraction reaches $\sim 50\%$ at $\sim 3\times 10^{12}\Msun$, 
while} nearly all satellites in the inner halo are quenched
above $\Mh \sim 10^{13}~\Msun$.  The difference may reflect the long duration of halo
quenching (more on this in \sec{discussion}).

Just as for the centrals, these results are evidence for two modes of
quenching, one related to $\sigone$ and one related to the halo, which
influences satellites differently in different regions of the halo.
To further understand what roles the halo and galaxy structure play in
quenching, we show the mean sSFR for 
satellites in \fig{morphmhdbinsgoodpsf} (bottom)
as a function of $\sigone$ and $\Mh$, divided in the same bins of
group-centric distance.

\fig{morphmhdbinsgoodpsf} (bottom) shows that while the behaviour of
sSFR is similar to $\fq$ in the outer and intermediate halo, the
decrease of sSFR with $\sigone$ is enhanced for satellites in the
inner regions of $\gtsima 10^{13}~\Msun$ haloes.  \sfc{Although all these
satellites are formally quenched (top panel), many appear to be near the
borderline.  For those, $\sigone$ is still able to modulate sSFR, as shown
in the bottom panel.}

\begin{figure}
\includegraphics[width=0.47\textwidth]{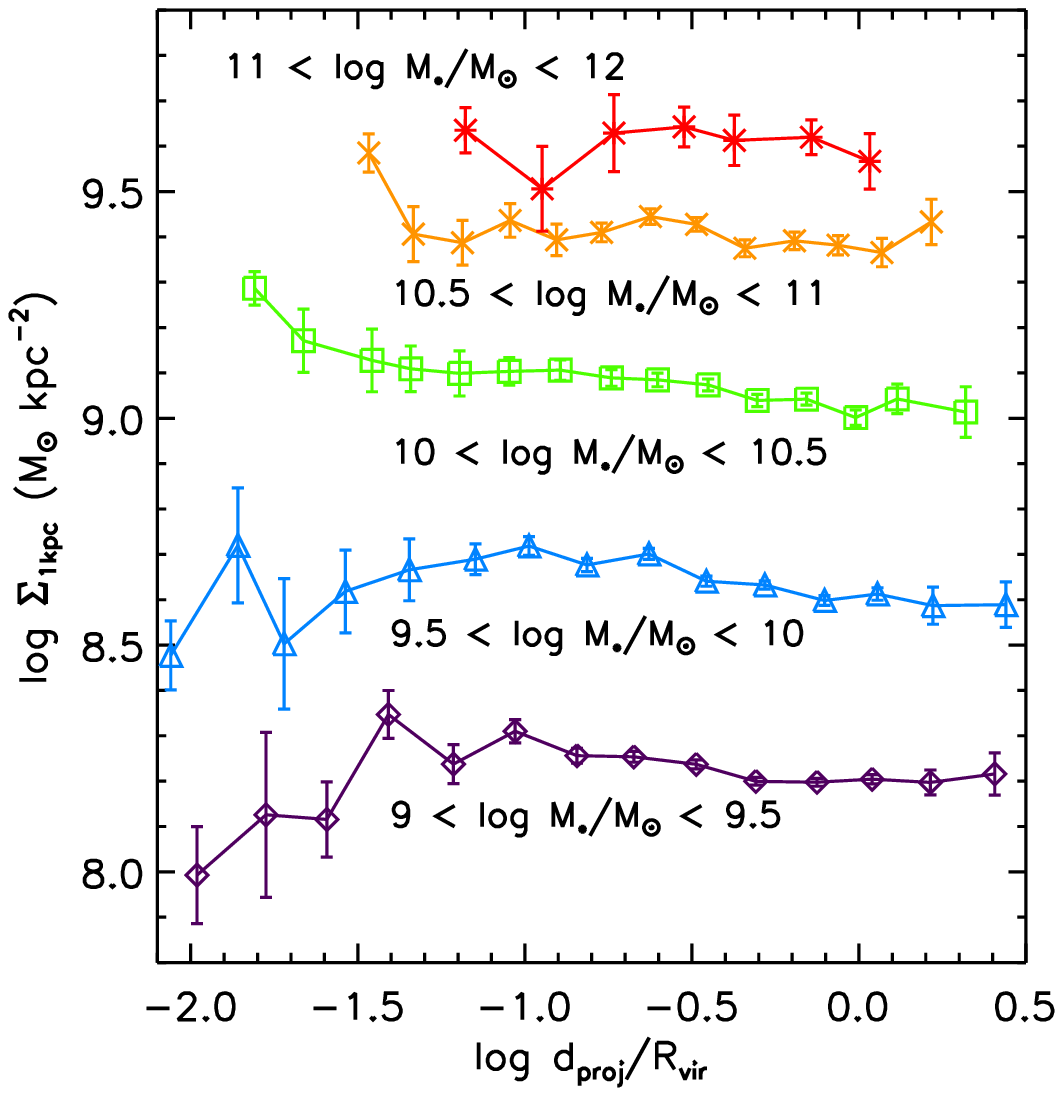}
\caption{\refc{Mean $\sigone$ as a function of $\Dist$ in bins of $\Ms$.  The 
error bars 
are errors on the means.  $\sigone$ varies only weaky with $\Dist$ for all 
masses, increasing at the most
by 0.1 dex between $\Rvir$ and $0.1\Rvir$. }}
\label{lsigvsdist}
\end{figure}

\refc{The environmental trends in \fig{morphmhdbinsgoodpsf} are independent of 
any relation 
between $\sigone$ and $\Dist$ because a horizontal strip in this figure is a 
strip of constant $\sigone$.  
In one such strip, say log $\sigone$ = 9, $\fq$ (and sSFR) above $\Mh = 10^{13} 
\Msun$ increases 
(decreases) toward the inner halo.  However, we show explicitly in 
\fig{lsigvsdist} that $\sigone$ varies 
only weakly with distance.  This plot shows $\sigone$ as a function of $\Dist$ 
in bins of $\Ms$.  
Clearly, $\sigone$ varies only weakly with $\Dist$ for all masses.  At the 
most, $\sigone$ increases 
by about 0.1 dex between $\Rvir$ and $0.1\Rvir$ which is smaller than the size 
of one of the pixels in 
\fig{morphmhdbinsgoodpsf}.  }

\begin{figure*}
\includegraphics[width=0.75\textwidth]{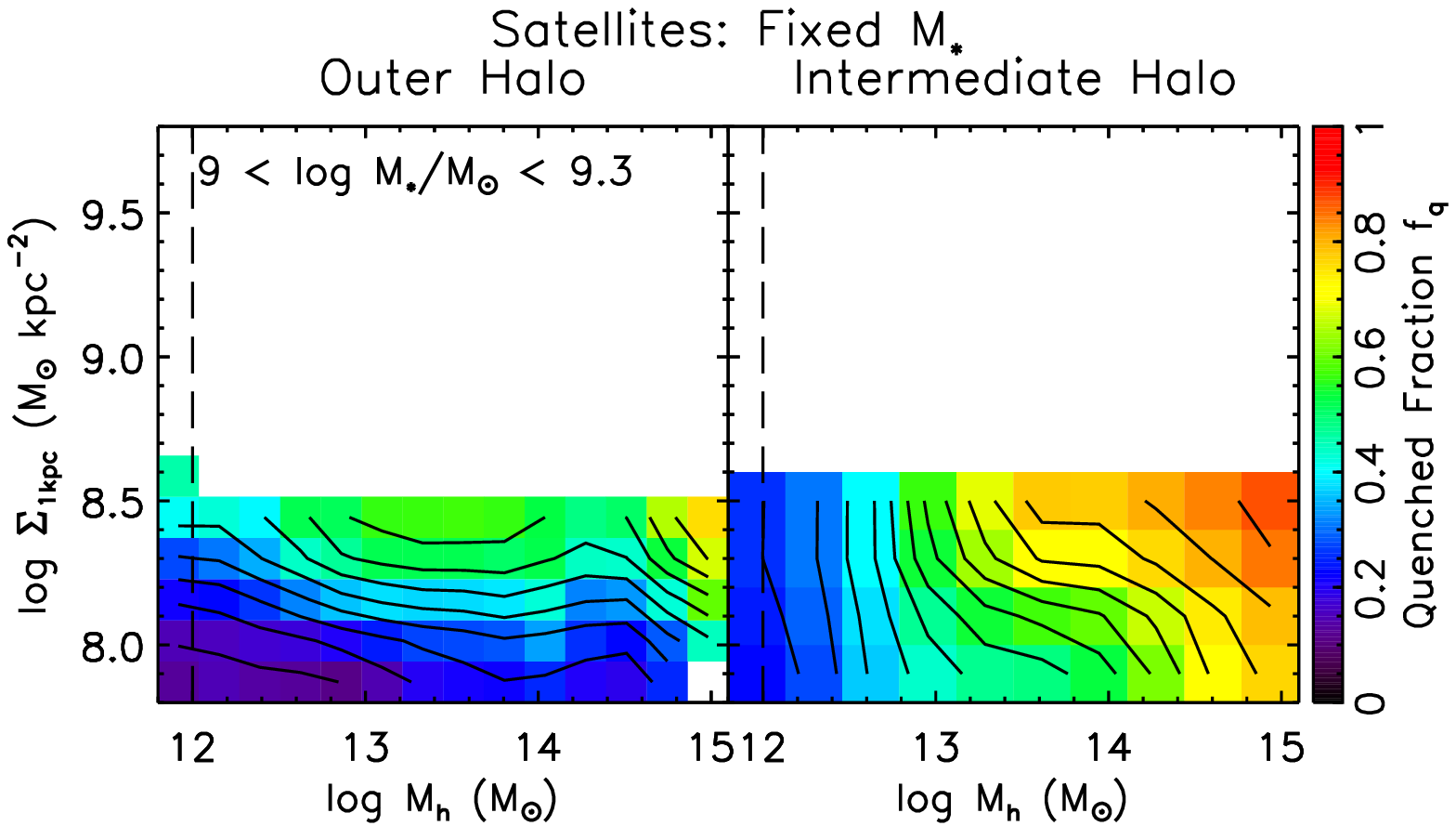}
\includegraphics[width=0.75\textwidth]{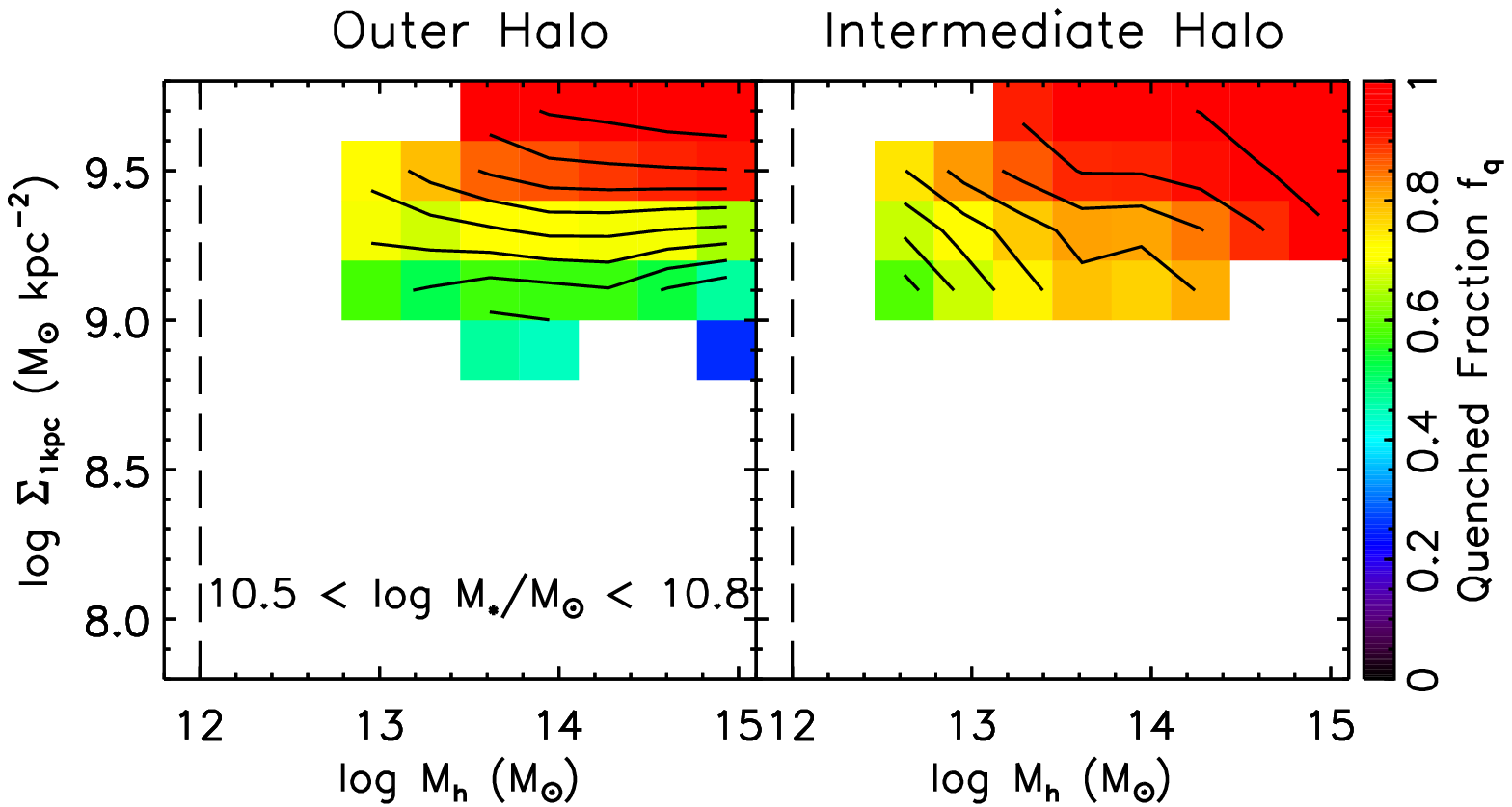}
\caption{\refc{The quenched fraction (top) and sSFR (bottom) of satellites
  as a function of the $\sigone$-$\Mh$ plane in two bins of
  group-centric distance and two bins of $\Ms$.  The ``Outer Halo'' refers to 
$\log (\Dist)
  > -0.5$, and the ``Intermediate Halo'' refers to $-1.0 < \log (\Dist) <
  -0.5$. This
  sample is limited to only those satellites for which 1 kpc is
  greater than the PSF width. 
  The black contours follow the colour scale and are 0.05
  apart for $\fq$ and 0.12 dex in yr$^{-1}$ for sSFR.  The vertical
  dashed line marks $\Mh = 10^{12}~\Msun$ below which the errors in
  the $\Mh$ estimates increase dramatically.  The steepening of the $\fq$-$\Mh$ 
relation
  in the intermediate halo (relative to the outer halo) is stronger for less 
massive satellites.}}
\label{satmassbins}
\end{figure*}

\refc{\fig{morphmhdbinsgoodpsf} includes satellites of all masses.  
We investigate quenching in the $\sigone$-$\Mh$ plane 
for satellites in two narrow bins of $\Ms$ in \fig{satmassbins}.   These
mass bins turn out to be two horizontal slices of \fig{morphmhdbinsgoodpsf} since 
$\sigone$ and $\Ms$ are strongly correlated.  The coloured shading shows $\fq$,
but sSFR is very similar.  These narrow slices of $\Ms$ contain too
few satellites with $\log (\Dist) < -1.0$ (the inner halo) to make any meaningful conclusions, 
so we omit
this distance bin in \fig{satmassbins}.  The black contours in both $\Ms$ bins 
are nearly horizontal
in the outer halo and steepen in the intermediate halo.  The difference in the 
quenching contours between the different regions of the halo is
greater for less massive satellites, but even massive galaxies 
feel the effects of the halo at intermediate distances.}

These results taken together show that halo-related quenching is
important in determining $\fq$ for satellites in the inner regions of
haloes.  Quenching that correlates with galaxy structure, which is most 
important
in the outer halo, also plays a role in the inner halo in determining
the satellites' mean sSFR.  We discuss a possible interpretation of these
results in \sec{discussion}.

\section{Discussion}
\label{discussion}

\subsection{Two Types of Quenching}

Our results suggest that both the halo and inner galaxy compactness play a
role in the quenching of central and satellite galaxies.  For
centrals, $\fq$ increases strongly with $\sigone$ while the halo
shifts the entire sSFR distribution.  Most quenched centrals are both
compact and have massive haloes, but one in five quenched centrals
above $\Mh = 10^{11.8}\Msun$ are diffuse ($\sigone <10^{9}\Msun
\kpc^{-2}$), indicating that they were not quenched via
$\sigone$-related processes.  For satellites $\sigone$ dominates
quenching on the outskirts of halos while the halo dominates satellite
quenching near the halo centre.

These findings are consistent with other observational work confirming
the importance of the bulge/central compactness in the quenching of galaxies
\citep{bell08,wuyts11,cheung12,bell12,barro13,omand14} and the
importance of the halo in quenching \citep{weinmann06,woo13,tal14}.
\cite{fang13} also find that both $\sigone$ and stellar mass $\Ms$ of
central galaxies are important predictors of quenching.  Since $\Ms$
for centrals is a only crude proxy for the halo mass, we have added to the
discussion by examining the role of the halo directly, as well as comparing
the quenching roles of mass and compactness for satellites.  Our results
are also consistent with those of \cite{bluck14}.  \arxc{They show that $\fq$ 
increases
with bulge mass (which is correlated with $\sigone$) at constant
$\Mh$ (their Fig. 11), and with $\Mh$ at constant bulge mass 
(but not with $\Ms$ - their Fig. 13).  Note that they do not point 
out the latter trend 
of $\fq$ with $\Mh$ which is clearly visible in their Fig. 13, 
but emphasize that the former trend is strongest}.

How shall we interpret these results, and in particular the different
behaviours of $\fq$ and sSFR?  We propose that the increase of $\fq$
is related to the transfer of galaxies from one side of the galaxy
bimodality to the other since it measures the fraction of galaxies
that are on one side.  The timescale for this transfer must be short
\sfc{so as not to fill up the} ``green valley'' (see
\fig{lssfrhists_fixedmh}).  In contrast, the decrease of sSFR refers to a
slower fading of star formation.  
Thus the quenching processes that
are related to compactness are quick while the halo process
is slow.\footnote{\sfc{Some may consider only the former as 
``quenching,'' and the latter more as star-formation regulation.
However, since the halo mechanism is thought to shut down gas accretion which would
otherwise continue to feed star formation, we will continue to refer to this
also as ``quenching'' in this discussion.}} 

Halo quenching is expected to operate in haloes of masses above a
threshold mass $M_{\rm crit}$ of order $10^{12}\msun$, where the
cooling time is longer than the relevant dynamical time. This enables
a stable shock at the virial radius, behind which the gas heats to the
virial temperature, such that gas supply to the central galaxy shuts
off \citep{bir03,dekbir06}. While the virial shock heating in a given
halo could be quick, $M_{\rm crit}$ is expected to vary by an order of
magnitude between different haloes because of their different
histories. For example, $M_{\rm crit}$ depends on metallicity, which
varies from halo to halo (\citealp{dekbir06}, Fig. 2). Furthermore,
once the halo mass is in the vicinity of $\sim M_{\rm crit}$, our
simulations demonstrate that in many cases the development of a virial
shock awaits a trigger, e.g., by a minor merger.  Indeed, simulations
show that the hot gas fraction is increasing very gradually with halo
mass, growing from $\ll 1$ to $\sim 1$ over almost two orders of
magnitudes in $M_{\rm h}$ about $M_{\rm crit}$
(\citealp{ker05,birnboim07,ocv08,keres09,vandevoort11}).  This
predicted behavior of halo quenching is consistent with the weak
dependence of $\fq$ and sSFR on $\Mh$ (at fixed $\sigone$) for
centrals.

Once all the halo gas is heated, cold gas that is already present
within the galaxy\footnote{Cold streams at $z>1-2$ may still bring gas
  into the most massive galaxies even in shock-heated halos
  \citep{dekel09}} is expected to continue forming stars until all the
gas is consumed or lost by feedback-driven outflows.  The typical gas
depletion time for massive galaxies is $\sim 2$-$3$ Gyr \sfc{(or longer for
early-type Sa disks)} in the local
universe \citep{pflammaltenburg09,bigiel11,saintonge11}, and at $z\sim
1$-$2$ \citep{saintonge11,dekel13,dekel14b}.  Therefore we expect the
shut down of the gas supply to be manifested in a slow fading of star
formation, or a decrease in sSFR.  This is consistent with the
decrease in sSFR with $\Mh$ for central galaxies.  This slower fading
produces a continous, rather than bimodal, distribution of sSFR as
seen in the absence of $\sigone$-related quenching
(red histogram of \fig{lssfrhists_fixedsigma}$a$).

On the other hand, bulge-building/compacting mechanisms that may
result in quenching, such as major mergers and gaseous inflows through
disc instability, \sfc{especially at high-$z$}, are inherently violent.  The 
resulting starbursts
lower the gas depletion timescales in ongoing mergers (and probably
also during violent disc instability) by factors of four to $>10$
\citep{young86,sanders91,sanders96,gao04}.  Powerful outflows due to
the stellar winds from these starbursts or from AGN fed by the gasous
inflows may further decrease the depletion time.  Thus we expect
quenching through these mechanisms to occur quickly compared to halo
shock heating.  Their quick nature may also imply that most galaxies
that quench through these processes quench at high-$z$
\citep{dekel14}.  Indeed, in hydrodynamical zoom-in simulations,
\cite{zolotov14} find that quenching driven by
compaction via instabilities takes less than a 1 Gyr.  Quick
bulge/compaction-related quenching is consistent with our observation
that $\fq$ increases primarily with $\sigone$ while sSFR for the
star-forming and quenched galaxies does not.  \sfc{While such processes seem to 
be rare at $z=0$
\citep{yesuf14}, many quenched galaxies observed today with high $\sigone$
may have quenched at high-$z$.}

However, once gas in consumed in a bulge-building/compacting starburst
and associated outflows, there must be a mechanism for preventing new
cold gas from accreting onto the galaxy in order to maintain its
quenched state.  The halo may play this role.  \sfc{Thus the overall 
``ripeness'' for quenching may 
be set by the halo while the moment of (quick) transition is triggered by 
internal, $\sigone$-related
processes.  }

Our result that
galaxies with high $\sigone$ in halos less massive than $10^{12}
\Msun$ have lower $\fq$ and higher sSFR than those in more massive
haloes (upper left corner of both panels of \fig{morphmass}) may be evidence that $\sigone$-related
processes are not enough to quench galaxies, and that the halo is
needed for quenching maintenance.  Indeed, most quenched centrals
are both compact and in massive haloes.  Our finding that
sSFR decreases with $\Mh$ for high-$\sigone$ galaxies may also point to an
increased efficiency of ``radio mode'' AGN feedback in hotter, more
massive haloes (\citealp{kormendy13} and references therein).
Galaxies with low $\sigone$ are not undergoing quick structure-related
quenching, and so these galaxies are an opportunity to observe the
slower $\Mh$-dependent quenching by itself.  Indeed for these, we
observe that both sSFR and $\fq$ correlate with $\Mh$.  Thus, for
those galaxies which experience both quenching mechanisms, the quicker
compaction-related processes may play the role of triggering quenching
while the slower halo process plays the maintenance role.

\refc{Since the majority of quenched centrals are both compact and in massive haloes, it may
also be possible that the dominant quenching mode is a single process related to both compactness and massive
haloes, rather than two independent channels with two different timescales.  After all, the
two are strongly correlated.  However, this work demonstrates that the halo and compactness work independently.  
The evidence for their independence is in the observed quenching of centrals in massive haloes that are diffuse, and 
of compact centrals that are in low mass haloes.  
Thus, rather than postulating a third (more dominant) channel where the halo and compactness are linked, we have preferred
the simpler scenario that these two independent channels work best together.  Distinguishing between these scenarios is
beyond the scope of this work.
}

\sfc{Not addressed by this analysis are the galaxies that seem to be quenching 
slowly and 
also building a dense bulge (slowly, e.g., pseudobulges).  
These bulges may only be incidentally growing while their star
formation is slowly fading due to the halo.  A study of the halo 
and bulge properties (classical vs. pseudobulges) of galaxies moving 
slowly through the green 
valley is needed to answer this question.}

\sfc{Also important is the fact that a significant area of the $\sigone$-$\Mh$ 
plane is strongly 
bimodal in sSFR.  This means that the combination of $\sigone$ and $\Mh$ does 
not perfectly predict quenching, 
and there remains at least one other unknown quantity important for quenching.  
This is related to the phenomenon 
of ``galactic conformity'' \citep{weinmann06,hartley14,knobel14} which states, among 
other things, that $\Mh$ does fully describe 
the quenched state of galaxies in a halo.  Quenching is correlated with 
$\sigone$ and $\Mh$, but possible 
time delays in quenching make the
physical mechanisms very hard to track.}

\sfc{Despite these uncertainties, however, the picture of quick 
$\sigone$-related
quenching and the halo regulating the slower fading of star formation seems to 
be a good first-order fit to 
our observations.  }

\begin{figure*}
\epsscale{2.0}
\plottwo{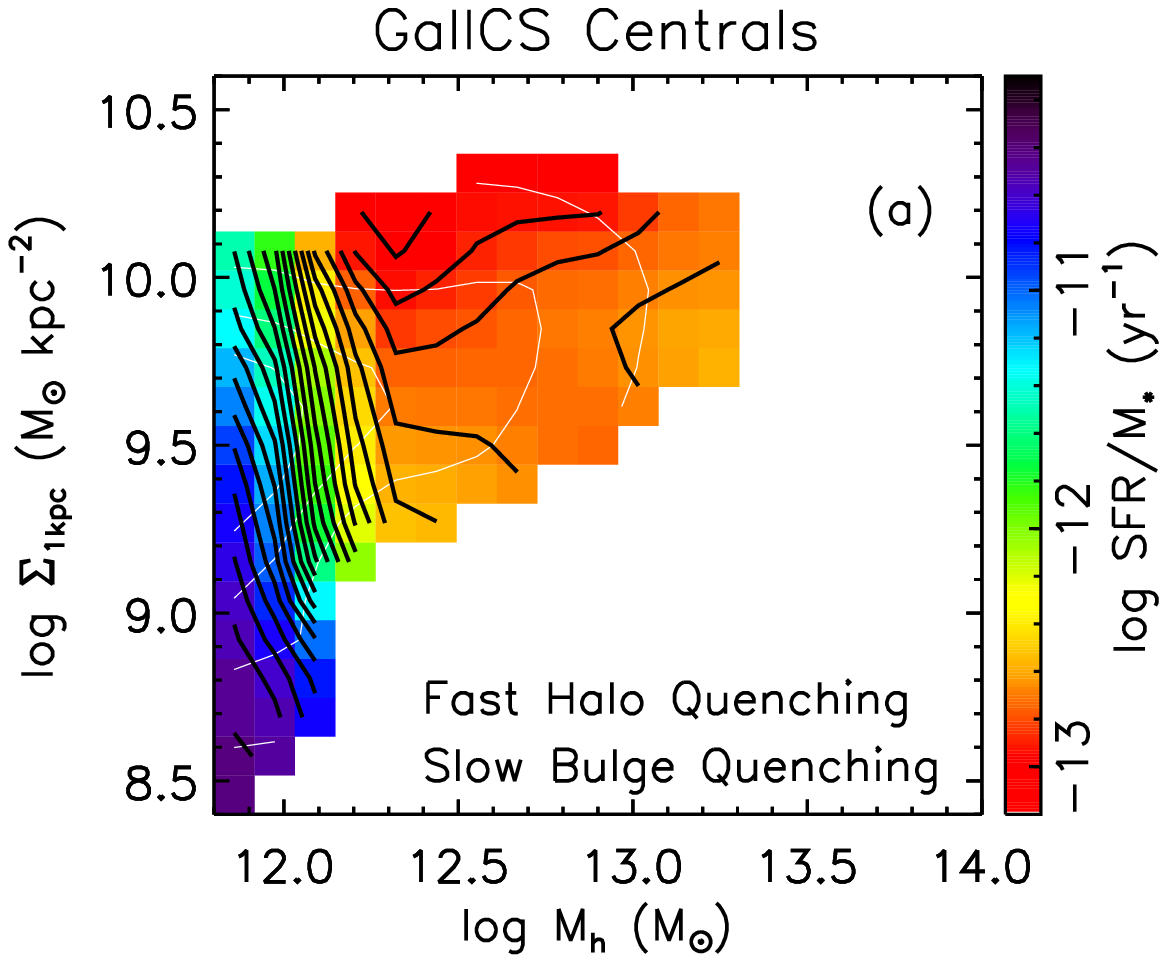}{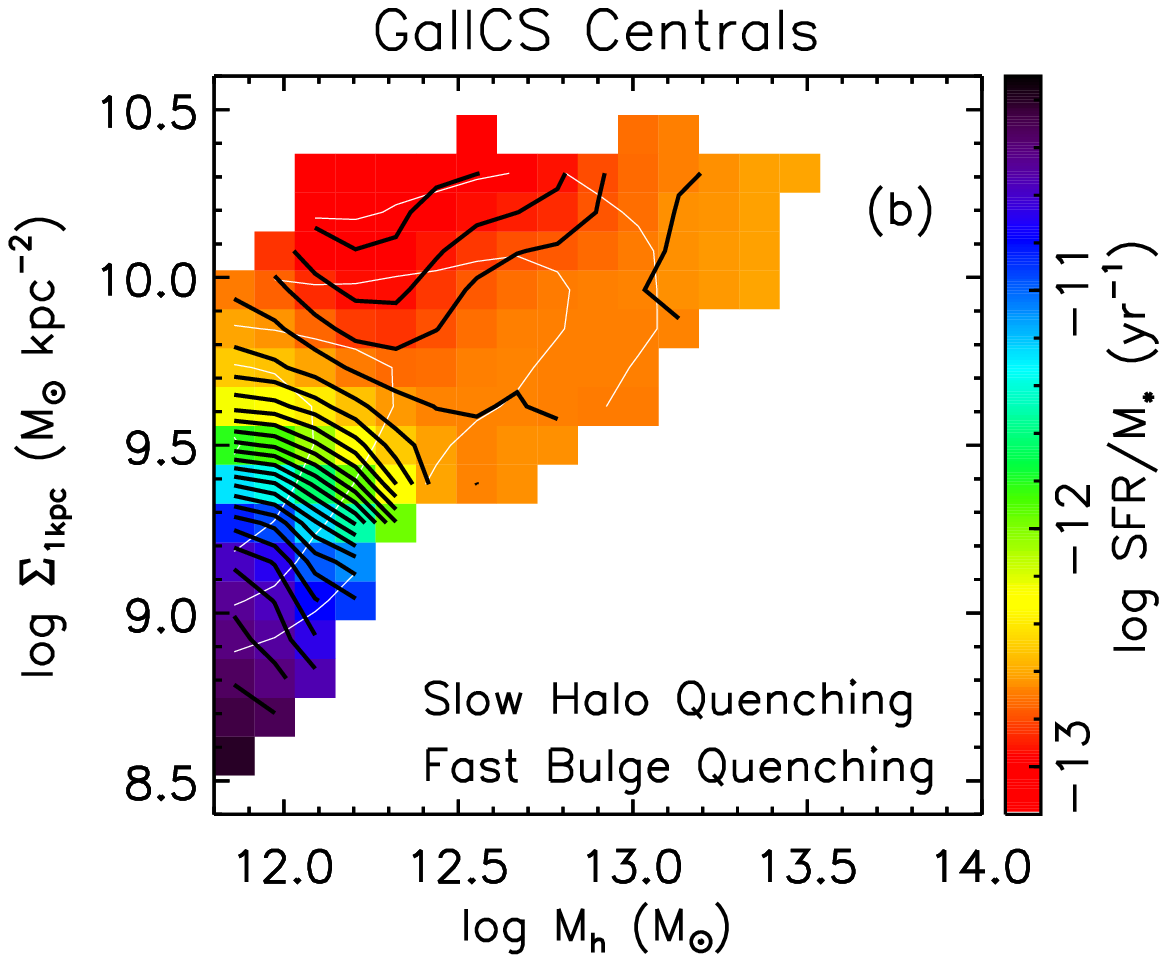}
\caption{Mean log SFR/$\Ms$ as a function of $\Mh$ and $\sigone$ in
  the GalICS semi-analytic model at $z=0$.  Quenching in 
  panel $a$ is implemented such that halo quenching cuts off accretion and
  removes gas from the central galaxy while bulge quenching only cuts
  off accretion.  In panel $b$, quenching is implemented such
  that halo quenching only cuts off accretion while bulge quenching
  both cuts off accretion and removes gas from the central galaxy.
  The white contours represent the number density of galaxies per
  pixel and are separated by 0.25 dex in Mpc$^{-3}$.  The black
  contours follow the colour scale and are 0.12
  dex apart in yr$^{-1}$.  The contours of sSFR in panel $b$
  are qualitatively more similar to \fig{morphmass}$b$ than 
  panel $a$.  }
\label{galics}
\end{figure*}

\subsection{A SAM Test}

The interpretation that the different durations of halo-related
quenching and compactness/bulge-related quenching mechanisms are
responsible for the observed quenching trends for centrals can be
tested via semi-analytic modelling.  We performed a preliminary
analysis of the GalICS semi-analytic model (SAM)
\citep{hatton03,cattaneo07,cattaneo08,cattaneo13} to test this and the
results are encouraging.  

\refc{The GalICS SAM runs on $N$-body merger trees  
of dark matter.  The baryonic prescription which populates the haloes with 
galaxies
and evolves them is nearly identical to that
of \cite{cattaneo13}.  In brief, gas is cooled onto discs whose sizes are 
determined
by conserving angular momentum.  Star formation is activated when the gas 
surface density reaches the Kennicutt threshold \citep{kennicutt89,ken98b}.  
Bulges are grown via mergers and 
disc instabilities which transfer mass from the disc to the bulge, passing 
through a 
starburst phase.  The amount of mass transfered is 
a function of the merger ratio, or the amount that
stabilises the disc in the case of a disc instability.  Stellar feedback 
returns mass and metals
to the cold medium.  Supernova energy returns cold gas to the hot medium or 
ejects gas depending 
on the deepness of the potential well.}

GalICS implements both halo-related
quenching and bulge-related quenching.  Halo-related quenching is
implemented by imposing a critical halo mass $\Mcrit$ (\refc{$= 10^{12}
\Msun$} and increases with $z$ above $z=3.2$ according to
\citealp{dekbir06}), above which accretion is cut off and all gas is
removed from the galaxy.  That is, halo quenching is implemented to be
immediate.  In contrast, for galaxies with bulge-to-total mass ratio
$B/T > 0.5$, only accretion is halted and remaining gas is allowed to
continue forming stars.  Thus, GalICS implements both halo- and
bulge-related quenching with durations that are the opposite of the
sense that we are suggesting from our SDSS results.  The resulting
sSFR as a function of $\sigone$ and $\Mh$ for central galaxies at
$z=0$ in GalICS is shown in \fig{galics}$a$\footnote{\small We computed
  $\sigone$ in GalICS by integrating the \cite{hernquist90} profile
  for the bulge and starburst components and an exponential for the
  disc component out to 1 kpc.  \refc{Galaxies for which $\sigone$ is low 
  despite having high $B/T$ (a rarity in the SDSS) are removed to ensure that 
any quenching 
  effect seen at low $\sigone$ is due to the halo alone.}}.  \refc{$\fq$ in this 
plane is similar.} 
  This combination of quenching
produces strong vertical contours of sSFR above $\Mcrit$ in contrast
to what is observed in the SDSS (\fig{morphmass}$b$).

\fig{galics}$b$ shows the prediction of the same model with the
sense of the quenching reversed, \ie, accretion is halted and gas is
removed from a galaxy when $B/T > 0.5$ while the $\Mcrit$ criterion
only cuts off accretion.  \refc{The plot of $\fq$ is similar.} 
Note that this reversal of the quenching is
the only change between the panels of \fig{galics}.  Qualitative
agreement with the SDSS is vastly improved.  Although quenching is
slightly too strong at high $\sigone$, the directions of the contours
of sSFR are in rough agreement with the SDSS (vertical for
low-$\sigone$ and roughly horizontal at mid-$\sigone$).  Given the large
uncertainties of the model, the qualitative agreement with the SDSS
after only one simple change is remarkable.  These results show that
two modes of quenching that effectively differ in duration can explain
much of the quenching trends with $\sigone$ and $\Mh$.  They at least
point the way for further study of quenching in SAMs, including
experimenting with gas accretion and ejection with different criteria
for halo- and compactness/bulge-quenching.

Thus, for central galaxies, galaxy compactness/bulges seems to play the role of
the quick transition to quiescence while the halo plays the role of
the slower fading of star formation.  This is consistent with the two
modes of quenching proposed in \cite{barro13,dekel14} consisting of an early
rapid quenching of compact star-forming galaxies and a later slower
quenching of more diffuse galaxies.  This is also consistent with the
results and interpretation of \cite{schawinski14} who also suggest
that quenching occurs in slow and fast modes.  These authors show that
early-type galaxies dominate the red sequence of galaxies whose
dust-corrected colour is largely unaffected by their halo mass.  On
the other hand, they show that the blue cloud and green valley are a
continuous population of slowly-evolving late-type galaxies whose
colour is strongly reddened above $\Mcrit$.  The late-type galaxies
above $\Mcrit$ in their analysis are likely dominated by satellites
(since they did not separate satellites from centrals), but we show
explicitly in our analysis that quenching trends for central galaxies
can be naturally explained if they experience both slow and fast modes
of quenching.

\subsection{Satellites}

This picture may also explain the quenching behaviour of satellites.
For these galaxies, group-centric distance is roughly an indicator of
how long they are influenced by their host halo.  Since halo quenching
is slower than $\sigone$-related quenching, $\fq$ and sSFR in the
outer halo correlate with $\sigone$ as they do for centrals and only
weakly with the host $\Mh$.  However, the influence of the halo here
is non-negligible, and increases in importance 
toward the inner halo
once the satellites have had enough time to be quenched by the halo.
\refc{This effect is strongest for less massive satellites.}
By this time, \sfc{$\fq$ has reached 50\% for satellites in the inner
  regions of haloes of mass $3\times 10^{12}\Msun$ ($\sim \Mcrit$).
  Above $\sim 10^{13}~\Msun$, these inner satellites} are all
quenched (\fig{morphmhdbinsgoodpsf}).  
Some of these quenched galaxies were in fact quenched
by $\sigone$ before they arrived in the inner halo.  These are the
ones with high $\sigone$, high $\Ms$, and low sSFR in the inner halo, and we
suggest that this is why sSFR decreases with $\sigone$ for quenched
satellites here.  

\refc{This interpretation is} consistent with the findings of
\cite{muzzin14} who study poststarburst satellites (\ie, those that
were quenched quickly) and find that these quench around 0.5R$_{200}$,
\ie, the left panels of \fig{morphmhdbinsgoodpsf}.  
This picture is also consistent
with findings of \cite{wetzel13} that satellite quenching timescales
are shorter at higher $\Ms$ but independent of $\Mh$, and with
\cite{wheeler14,taranu14} who find that low-mass satellites must have slow
quenching timescales, and at least slower than centrals \citep{tal14}.

\cite{wetzel13,trinh13,mok13} proposed that satellite quenching is a
``delayed-then-rapid'' process since neither rapid nor slow quenching
adequately explain their observations.  Instead, our framework
suggests that the satellite population experiences a combination of
\sfc{slow and rapid quenching processes rather than a strictly
  chronological sequence.  The quick process is the $\sigone$-related
  quenching in the outskirts of haloes that also occurs in centrals,
  and the slower quenching is $\Mh$-related that is finally manifest
  once the satellite reaches the inner halo.  Satellites at
  intermediate radii vary smoothly from one extreme to the other.}  

In addition, ram pressure stripping (a mechanism of halo quenching for
satellites) strips only the gaseous halo of satellites at $\sim 1
\Rvir$ from the halo centre.  As a satellite migrates inward, ram
pressure also strips cold gas, starting with the gas in the outer disc
(Zinger et al. in prep.) which may lead to more rapid quenching.

\section{Conclusion}
\label{conclusion}

In summary, our results include:
\begin{itemize}
\item Central stellar compactness strongly correlates with $\fq$ for
  centrals at fixed $\Mh$, especially for mid-range values of
  $\sigone$ ($\sim 10^{9-9.4}$).  $\Mh$ correlates with sSFR (and
  $\fq$ around $\Mcrit$) at fixed $\sigone$ for centrals with higher and
  lower $\sigone$.  
\item For central galaxies at fixed $\Mh$, the shape of the
  distribution of sSFR changes with $\sigone$ such that galaxies with
  higher $\sigone$ have a more numerous passive population.  However
  at fixed $\sigone$, increasing $\Mh$ shifts the entire distribution
  of sSFR to lower values without a significant change in shape.  
  \refc{This is true at also fixed $\Ms$.  
  However, varying $\Ms$ at fixed $\Mh$ and $\sigone$
  does not change the sSFR distribution.}
\item Most quenched centrals are both compact and live in massive
  haloes.  However one in five quenched centrals above $\Mh = 10^{11.8}\Msun$
  are diffuse ($\sigone < 10^{9}\Msun \kpc^{-2}$).  These may have
  been quenched by the halo alone, since they certainly did not quench
  through compaction-related processes.
\item $\Mh$-dependent quenching of satellites (at constant $\sigone$) is
  seen most strongly in the inner regions of haloes.  $\fq$ and sSFR
  correlate strongly with $\sigone$ and weakly with $\Mh$ in the outer
  regions of haloes.  The correlation of quenching with $\Mh$ becomes
  stronger, \refc{especially for less massive satellites,} toward the inner 
  halo.  \sfc{Here, $\fq \sim 0.5$ at $\sim \Mcrit$, with satellites almost completely
  quenched in haloes $\Mh \gtsima 10^{13}~\Msun$}.  sSFR decreases with $\sigone$ for
  these quenched satellites.
\end{itemize}

Our results suggest that both the halo and galaxy compactness play a
role in the quenching of central and satellite galaxies.  Galaxy inner
compactness determines $\fq$ while $\Mh$ determines sSFR for
star-forming and quenched centrals, perhaps
pointing to the quick and slower timescales of bulge/compactness- and
halo-related quenching as demonstrated in a SAM.  
For satellites, halo
quenching becomes manifest once they have reached the inner halo,
where nearly all are quenched above $\Mcrit$.  But along the way,
compactness-related quenching operates on satellites independent of the
halo.

\section*{Acknowledgements}
We thank the anonymous referee for helpful comments that improved this paper.
We acknowledge the helpful and stimulating discussions with Yuval
Birnboim, Marcella Carollo, Will Hartley, Simon Lilly, Nir Mandelker,
Kevin Schawinski, Benny Trakhtenbrot, Megan Urry and Andrew Wetzel.  We thank
Marcello Cacciato for kindly sharing his halo mass function code, and Andrea 
Cattaneo for assistence with the GalICS code.  We
also thank Frank van den Bosch and Xiaohu Yang for kindly providing
the group catalog for the SDSS DR7. This research has been partly
supported at HU by ISF grant 24/12, by GIF grant G-1052-104.7/2009, by
a DIP grant, by NSF grant AST-1010033, and by the I-CORE Program of
the PBC and the ISF grant 1829/12.
SMF and DCK acknowledge partial support for this work from an
NSF grant AST 08-08133.
Funding for the SDSS and SDSS-II has been provided by the Alfred P. Sloan Foundation, the Participating Institutions, the National Science Foundation, the U.S. Department of Energy, the National Aeronautics and Space Administration, the Japanese Monbukagakusho, the Max Planck Society, and the Higher Education Funding Council for England. The SDSS Web Site is http://www.sdss.org/. The SDSS is managed by the Astrophysical Research Consortium for the Participating Institutions. The Participating Institutions are the American Museum of Natural History, Astrophysical Institute Potsdam, University of Basel, University of Cambridge, Case Western Reserve University, University of Chicago, Drexel University, Fermilab, the Institute for Advanced Study, the Japan Participation Group, Johns Hopkins University, the Joint Institute for Nuclear Astrophysics, the Kavli Institute for Particle Astrophysics and Cosmology, the Korean Scientist Group, the Chinese Academy of Sciences (LAMOST), Los Alamos National Laboratory, the Max-Planck-Institute for Astronomy (MPIA), the Max-Planck-Institute for Astrophysics (MPA), New Mexico State University, Ohio State University, University of Pittsburgh, University of Portsmouth, Princeton University, the United States Naval Observatory, and the University of Washington.

\bibliographystyle{mn2e}
\bibliography{jobib}

 \label{lastpage}

 \end{document}